\documentclass[12pt,preprint,apj]{emulateapj}

\newcommand{\kms}{{km s$^{-1}$}}
\newcommand{\kmsMpc}{{km s$^{-1}$ Mpc$^{-1}$}}

\usepackage{epstopdf}
\usepackage{multirow}
\usepackage{color}

\shorttitle{Star Clusters in NGC 4921}
\shortauthors{Lee \& Jang}

\begin{document}

\title{ 
Globular Clusters and Spur Clusters in NGC 4921, the Brightest Spiral Galaxy in the Coma Cluster}

\author{  Myung Gyoon Lee \& In Sung Jang}
\affil{Astronomy Program, Department of Physics and Astronomy, Seoul National University, Gwanak-gu, Seoul 151-742, Korea}
\email{mglee@astro.snu.ac.kr,  isjang@astro.snu.ac.kr}


\begin{abstract}

We resolve a significant fraction of globular clusters (GCs) in NGC 4921, the brightest spiral galaxy in Coma. 
Also we find a number of extended bright star clusters (star complexes) in the spur region of the arms. The latter are much brighter and bluer than those in the normal star-forming region, being as massive as $3 \times 10^5 M_\odot$. 
The color distribution of the GCs in this galaxy is found to be bimodal.
The turnover magnitudes of the luminosity functions (LF) of the blue (metal-poor) GCs ($0.70<(V-I)\leq1.05$) in the halo are estimated to be $V({\rm max}) =27.11\pm0.09$ mag and $I({\rm max})=26.21\pm0.11$ mag. 
We obtain similar values for NGC 4923, a companion S0 galaxy,
and two Coma cD galaxies (NGC 4874 and NGC 4889). 
The mean value for the turnover magnitudes of these four galaxies
is $I$({\rm max})$=26.25\pm0.03$ mag.
Adopting  
$M_I ({\rm max}) = -8.56\pm0.09$ mag for the metal-poor GCs, we determine the mean distance to the four Coma galaxies, $91\pm4$ Mpc. 
Combining this and the Coma radial velocity,
we derive a value of the Hubble constant, $H_0 = 77.9\pm3.6$ \kmsMpc.
We estimate
the GC specific frequency of NGC 4921 to be
$S_N = 1.29\pm0.25$, close to the values for early-type galaxies. This indicates that NGC 4921 is in the transition phase to S0s.

\end{abstract}

\keywords{galaxies: spiral  --- galaxies: clusters: individual (Coma, NGC 4921, NGC 4923, NGC 4874, and NGC 4889)   --- galaxies: star clusters: general  --- galaxies: distances and redshifts} 

\section{INTRODUCTION}

NGC 4921 is the brightest 
spiral galaxy (SB(rs)ab) in the Coma cluster \citep{dev91}.
It is only 0.5 mag fainter than NGC 4889, a cD galaxy in Coma,
and is as bright as M87, a cD galaxy in Virgo.
It hosts an AGN, and has an S0 companion, NGC 4923 ((R)SA(r)0-?), at 2\farcm6  southeast of NGC 4921.
{\bf Table \ref{tab_param}} 
lists the basic information of NGC 4921 and NGC 4923.
The heliocentric radial velocity of NGC 4921, 5482 \kms, is somewhat smaller than that of the Coma cluster, 6925 \kms \citep{str99}. Thus it has been considered that NGC 4921 is infalling to the Coma cluster \citep{vik97,bro11,and13}.

NGC 4921 is, as well as NGC 4569 (M90) in Virgo, a prototype of anemic spiral galaxies introduced in the revised DDO classification system \citep{van76}.
Anemic spiral galaxies have smooth spiral arms whose surface brightnesses are much fainter than those of normal spiral galaxies. They contain little gas. These facts indicate that  star formation activity in these galaxies is much lower than those in normal spiral galaxies with the similar subclass. 
Since these anemic galaxies are found in galaxy clusters like Virgo and Coma, it has been considered that anemic galaxies lost a significant fraction of gas due to the environmental effects such as ram pressure stripping and tidal stripping 
 (see \citet{van76,vik97} and \citet{ken15}).
\citet{van76} noted that NGC 4921 has an extremely red color among the spiral galaxies with similar luminosity in Virgo,
and pointed out that gas depletion was more severe in Coma than in Virgo. 
Recently \citet{ken15} presented a strong evidence based on the HI map suggesting  that ram pressure is acting on the gas in this galaxy.

 Modern wide field surveys found
that the color-magnitude diagram of nearby galaxies shows two major populations, a red sequence and a blue cloud, which are connected by a green valley (\citet{bla09} and references therein). 
Recent studies using a large number of galaxies suggest that
red spiral galaxies are in the transition phase from the blue cloud to the red sequence  (\citet{toj13,sme15,lee15} and references therein). 
NGC 4921, because of its proximity to us, provides a snapshot showing the details of this transition phase of galaxy evolution.

\begin{deluxetable*}{lccc}
\setlength{\tabcolsep}{0.05in}
\tablecaption{Basic Parameters of NGC 4921 and NGC 4923}
\tablewidth{0pt}
\tablehead{ \colhead{Parameter} & \colhead{NGC 4921} &\colhead{NGC 4923} & \colhead{Reference}}
\startdata
R.A.(2000) 	& $13^h01^m26.^s1$ &		$13^h01^m31.^s8$	& RC3	\\
Dec(2000) 	& $27\arcdeg53\arcmin10\arcsec$& $27\arcdeg50\arcmin51\arcsec$&RC3	\\
Type		& SB(rs)ab		& (R)SA(r)0				& RC3	\\
Distance moduli, $(m-M)_0$& $34.73\pm0.14$ & 	 $34.93\pm0.26$ & This study\\
Distance , d [Mpc] 	& $88.3\pm5.8$ & 	 $96.8\pm11.4$ & This study\\
Foreground extinction, $A_B$, $A_V$, and $A_I$ & 0.035, 0.027, and 0.015 &0.033, 0.025, 0.014 & \citet{sch11} \\
Total magnitude, $B^T$	& $13.04\pm0.15$ & $14.67\pm0.13$ & RC3 \\
Total color, $B^T-V^T$	& $0.87\pm0.02$ & $0.95\pm0.01$& RC3 \\
Ellipticity, $e=(a-b)/a$ & $0.11\pm0.09$ & $0.00\pm0.16$ & RC3 \\
$B$ absolute magnitude, 	& $M_B = -21.72\pm0.21$ & $M_B = -20.29\pm0.29$, & RC3, This study \\
$V$ absolute magnitude, 	& $M_V = -22.59\pm0.21$ & $M_V = -21.24\pm0.29$, & RC3, This study \\
Position angle			& 164 deg (B), 135 deg (K)  & No data in RC3 & RC3 \\
$R_{25} (B)$				& $75\farcs0 \times 60\farcs0$ & $34\farcs49 \times 26\farcs9$ & NED \\
$R_{total}(K)$				& $105\farcs2 \times 84\farcs2$ & $32\farcs2 \times 23\farcs83$  & NED \\
Heliocentric velocity, $v_h$ & $5482\pm4$ \kms & $5484\pm8$ \kms &RC3, NED \\
\hline
\enddata
\label{tab_param}
\end{deluxetable*}

Although NGC 4921 has anemic arms, it shows a number of impressive spurs (or outgrowths) associated with bright star clusters in its western arms. \citet{car13} and \citet{ken15} studied the morphology of these spurs, but suggested different scenarios to explain the origin of these spurs. 
\citet{tik11} presented $VI$ photometry of star clusters in NGC 4921 and NGC 4923, 
derived from the Hubble Space Telescope (HST)/Advanced Camera for Surveys (ACS) archival images. 
They found several thousand GC candidates in NGC 4921, which is an order of magnitude more than the number of GCs found in normal spiral galaxies like the Milky Way Galaxy and M31. They noted that the color distributions of the GCs in NGC 4921 and in NGC 4923 are bimidal. From the luminosity functions of these GCs (GCLFs), they estimated a distance of $97\pm5$ Mpc to the pair of NGC 4921 and NGC 4923.

\citet{tik11} used only a fraction of the archival images of NGC 4921 and selected only point sources in the images for their study, assuming that all GCs in these galaxies are starlike in the ACS images.
However, the ACS images we produced as described in the next section have a higher spatial resolution than those used in \citet{tik11}, and they show a number of compact sources that appear as slightly extended sources. Most of these sources are probably GCs in NGC 4921 and in NGC 4923.
In addition these images show a number of young star clusters and star complexes that are more extended than the GCs, which were not studied in \citet{tik11}.

In this study we resolve the GCs and extended star clusters in NGC 4921 and NGC 4923, taking advantage of the high resolution ACS images produced via drizzling, and investigate their properties in detail.
The main goals of this study are as follows.
First, we investigate various properties of the subpopulations in the star cluster systems of NGC 4921 and NGC 4923. 
Second, we study the photometric properties of the star clusters forming in the spur region of NGC 4921 in comparison with those in the normal star-forming region to find any difference between the two.
Third, we derive the GCLFs for these galaxies, and use them to estimate the distances to these galaxies. In particular, we use the LF for blue (metal-poor) GCs for distance determination, while previous studies on Coma galaxies are based on mostly the LFs for the entire GC samples \citep{kav00,woo00,har09, tik11}.
The distance to the Coma cluster is not yet well-determined \citep{car08}. 
The rich GC systems in NGC 4921, NGC 4923, and Coma cD galaxies enable us to derive reliable distances to these galaxies, which is useful in determining a mean distance to Coma and a value of the Hubble Constant.

\begin{figure*}
\centering
\includegraphics[scale=1.4]{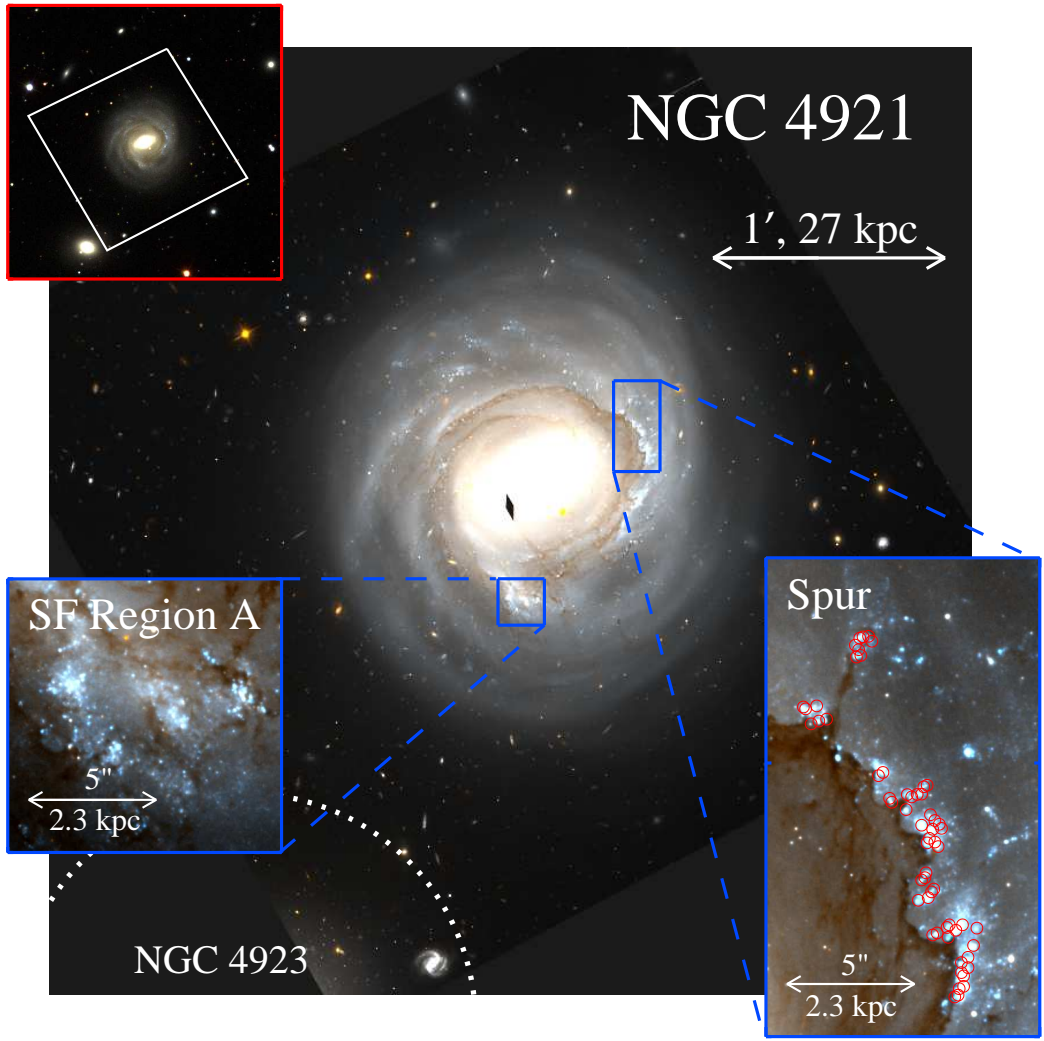} 
\caption{ Color images of NGC 4921. 
(Top left) Identification of stacked HST fields (solid solid line) overlayed on the $6\arcmin \times 6\arcmin$ SDSS image. The bright galaxy to the south-east is NGC 4923 (S0).
(Middle) A color image of the HST Field.  F814W  is red,  F606W  is green, and ($2 \times $F606W--F814W) is blue. North is up, east to the left.
(Bottom left) A zoom-in view of Star-forming Region A.
(Bottom right) A zoom-in view of the spur region where bright star clusters are lying in the boundary of dust lanes and spurs. Circles represent the detected sources along the dark lanes and spurs. Arrows denote the scales.}
\label{fig_finder}
\end{figure*}
 
This paper is composed as follows. In Section 2 we describe  the data we use and how we select star cluster candidates.
\S3 presents color-magnitude diagrams of the GCs 
in NGC 4921 and in NGC 4923, 
and young star clusters and star complexes in the spur region and star-forming region in NGC 4921. 
We show color distributions of the GCs in NGC 4921 and in NGC 4923. 
We derive GCLFs for these two galaxies.
Then we present the spatial  and radial distribution of the GCs in comparison with the surface brightness profiles of NGC 4921.
In \S4 we determine the distances to these galaxies using the GCLFs, and present similar results for two cD galaxies in Coma.  Using the mean distance to these four Coma galaxies, we measure the value of the Hubble Constant.
We estimate the specific frequency of the GCs in NGC 4921 and discuss it in context of morphological transformation.
Furthermore we discuss implications of the results on the spur clusters in relation with the origin of the spurs. 
Primary results are summarized in the final section.


\begin{figure}
\centering
\includegraphics[scale=0.9]{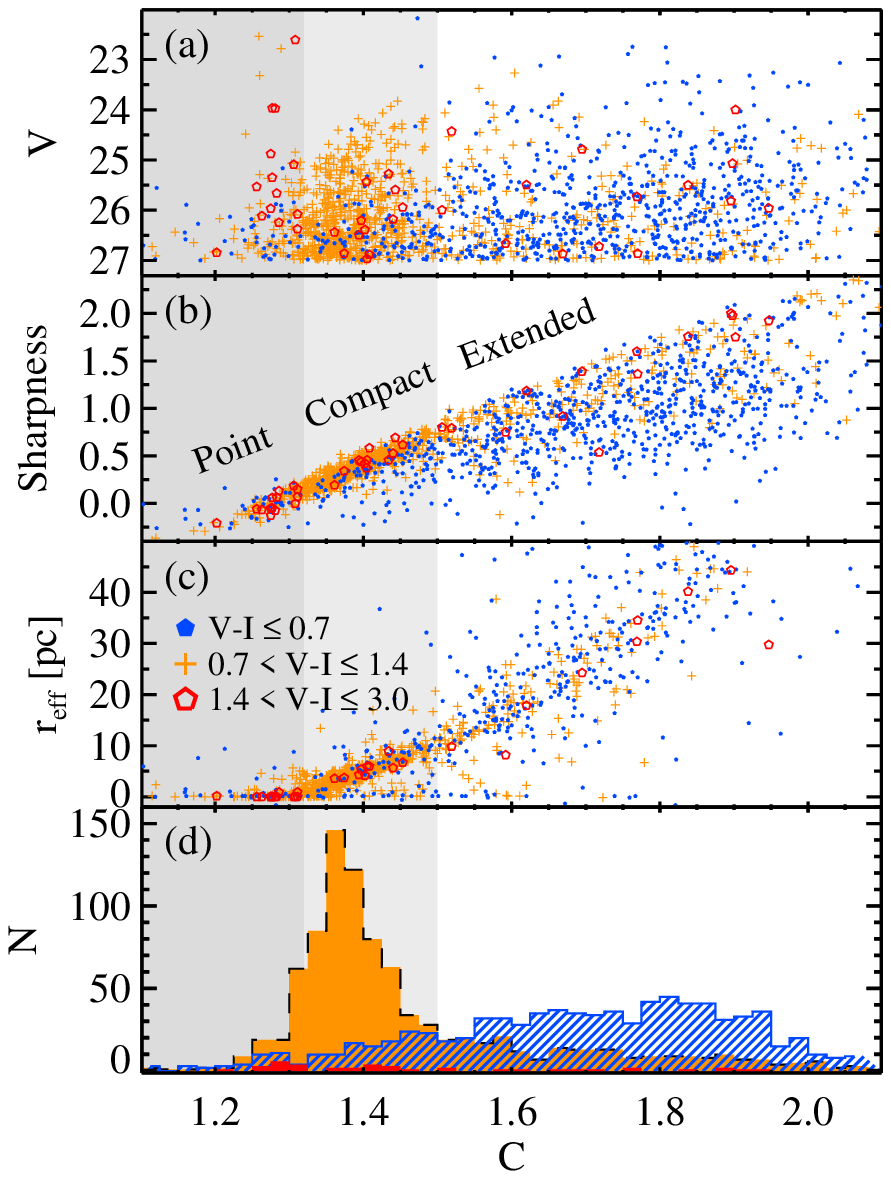} 
\caption{Concentration index ($C$) of the detected sources with $V<27$ mag measured from the $F814W$-band vs. (a) $V$-band magnitudes, (b) $F814W$-band sharpness, and (c) $F814W$-band effective radii. In each panel, filled pentagons, crosses, and open pentagons denote the blue sources with $(V-I)\leq0.7$, GC-like color sources with $0.7 < (V-I) \leq 1.4$, and red sources with $1.4 < (V-I) \leq 3.0$, respectively. 
The concentration index distributions for  the sources with different colors are represented by hatched, filled dashed-lined, and filled histograms, respectively, in panel (d).
Note that there is a strong concentration with a peak at $C\approx 1.37$, which is mainly composed of resolved GCs.
We divided the sources into three groups:
point sources ($C \leq 1.32$), compact sources ($1.32<C \leq 1.5$), and extended sources ($1.5<C \leq 2.2$).
We selected the sources with $C \leq 1.5$ as GC candidates.
}
\label{fig_CI}
\end{figure}

\section{DATA AND DATA REDUCTION}

We used  F606W and  F814W images of NGC 4921  (PID: 10842) from the HST archive. 
These images include part of a north-west region 
of NGC 4923. 
We then produced  master images of each filter using the STScI/drizzle package as described in \citet{jan14, jan15}.
We adopted a pixel scale of $0\farcs03$ (corresponding to 12.8 pc at the distance of 88 Mpc), and a pixfrac value of 0.7. {  The parameter pixfrac represents the ratio of the linear size of the finer pixel (called a drop) to the input pixel’s linear size before geometric distortion corrections.}
Total exposure times for the drizzled images are 62,240 s for  F606W  and 37,344 s for  F814W.

Note that \citet{tik11} used only 15 F606W images (with a total exposure time of 37,350 s) and 10 F814W images (with a total exposure time of 25,020 s),  because the remaining images were obtained in different spatial orientations. We used all the ACS images of NGC 4921 available in the archive, because the drizzle package can combine images obtained with different spatial orientations.
Thus, the images used in this study have higher spatial resolution and can reach deeper photometry than those used by \citet{tik11}.


{\bf Figure 1} displays a pseudo-color image of NGC 4921 we made by combining the images for each filter. The diffuse emission in the south-east region of the field represents the outer region of NGC 4923 whose center is $2\farcm6$ to the south-east direction from the NGC 4921 center.
Several spurs are seen along the dark dust lanes in the western region of the disk, and  a significant number of bright knots (which are star clusters or complexes) are found close to the spurs and dust lanes. We selected a spur region where several dark cloud spurs are seen to study these bright sources. In addition, we selected 
a normal star-forming region in the southern region of the disk, Star-forming Region A, for comparison,
as marked in Figure 1 and shown by the zoomed-in images.

We used DAOPHOT in IRAF to detect the sources in the images and obtained  their magnitudes with point-spread function (PSF) fitting \citep{ste94}. We adopted a 3$\sigma$ threshold for source detection. We transformed the instrumental magnitudes of the sources onto the
standard Johnson-Cousins $VI$ system in the Vega magnitude system using the information in \citet{sir05}. 
Finally, we carried out a careful visual inspection of the images of 7,068 detected sources with $V \leq 28$ mag, and removed 2,766 sources that appeared 
to be artifacts, background galaxies, or irregular structures. 
Only 4,302 sources with $V \leq 28$ mag were used in the following analysis related with colors.  Additional faint sources with  $28 < V \leq 29$ mag were  included for the analysis of the star clusters in the star-forming region and  the GCLFs.

\begin{figure*}
\centering
\includegraphics[scale=0.85]{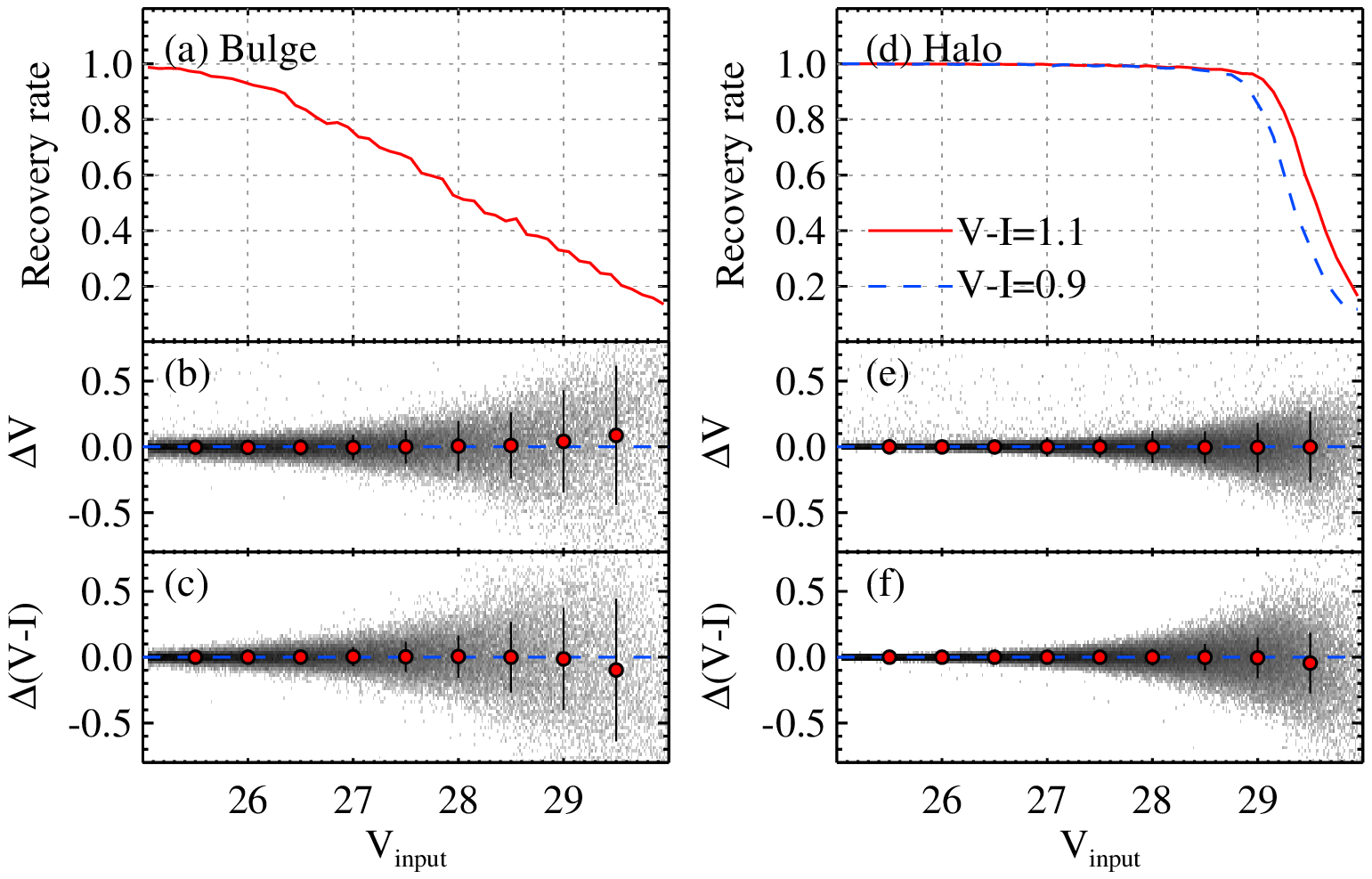}
\caption{Completeness of the photometry for the compact sources (with $C=1.37$) in the bulge and halo regions of NGC 4921.
(a) Recovery rates for the red ($(V-I)=1.1$) GC-like sources  in the bulge region.
(d) Recovery rates for the blue ($(V-I)=0.9$) and red GC-like sources in the halo region.
(b,c,e,f) Differences in $V$ and $(V-I)$ (inputs minus outputs) vs. input $V$ magnitudes for the bulge and halo regions.}
\label{fig_compl}
\end{figure*}

\begin{figure}
\centering
\includegraphics[scale=0.75]{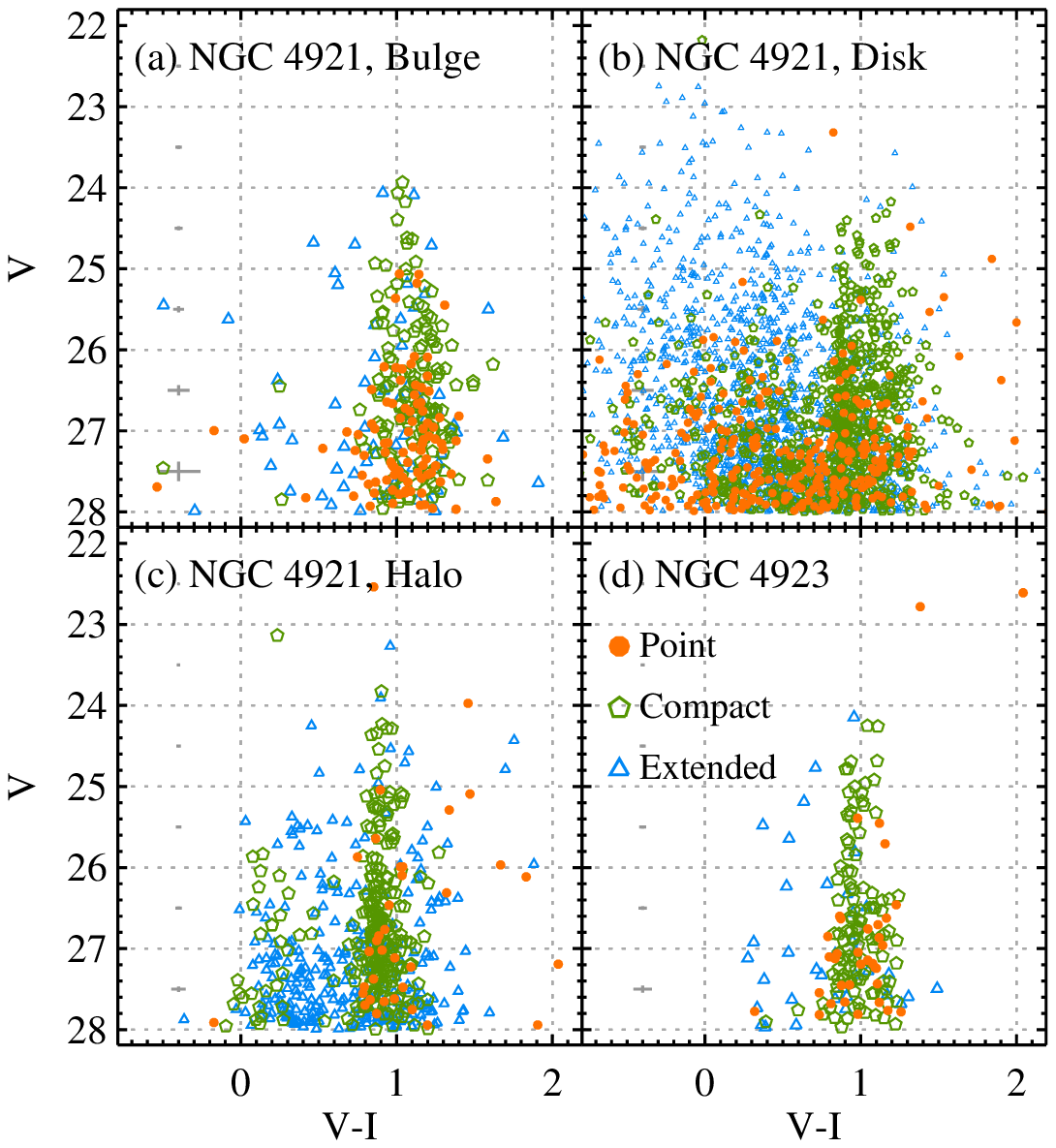} 
\caption{CMDs 
for the detected sources in the (a) bulge ($6\arcsec<R_{N4921}\leq17\arcsec$), (b) disk ($17\arcsec<R_{N4921}\leq70\arcsec$), and (c) halo ($R_{N4921}>70\arcsec$) of NGC 4921.  (d) CMD for the region ($R_{N4923}\leq1\arcmin$) of NGC 4923. Filled circles, open pentagons, and open triangles represent  point sources ($C\leq1.32$), compact sources (slightly extended sources, $1.32<C\leq1.5$), and extended sources ($1.5<C\leq2.2$), respectively. 
Note that compact sources are mostly GCs, and extended bright sources are mostly young star clusters and complexes in NGC 4921.}
\label{fig_cmd}
\end{figure}

\subsection{Photometry of Star Clusters} 

At the distance of Coma, GCs are expected to appear as point sources
in individual HST images. Considering this, \citet{tik11} selected star-like sources in their study of GCs in NGC 4921.
However, we noted that many GCs are slightly resolved in our master drizzled images. 
We measured the concentration index ($C$) of the detected sources, which is defined as the difference between the small and large aperture magnitudes (aperture radii of $0\farcs035$ and $0\farcs109$): $C = mag(r=0\farcs035) - mag(r=0\farcs109)$.
We measured the values of $C$ in the $F814W$-band image to avoid contamination by $H\alpha$ emissions in the star-forming regions  in the $F606W$-band image.
 
{\bf Figure \ref{fig_CI}} displays the distribution of the measured $C$ values of the detected sources with $V\leq27$ mag,
against $V$ magnitudes (a), DAOPHOT sharpness parameter measured in the $F814W$-band image (b), effective radii ($r_{\rm eff}$) (c), and their histograms (d).
{  We measured the effective radii of only the bright sources with $V\leq27$ mag using ISHAPE \citep{lar01}, avoiding large errors in the size estimations for fainter sources.}
In this figure, we plotted the sources with blue colors ($(V-I)\leq0.7$), GC-like colors ($0.7<(V-I)\leq1.4$), and red colors ($1.4<(V-I)\leq3.0$) using filled pentagons, crosses, and open pentagons, respectively.

The histogram of the sources with GC-like colors shows a strong peak at $C \sim 1.37$, and a long weak tail 
towards large values of $C$. 
On the other hand, the histogram of the sources with blue colors shows a much broader distribution, showing that most of the blue sources are larger than the GC-like sources.
The histogram of the sources with red colors shows two components: one narrow component with a peak at $C\sim1.28$ which consists of unresolved sources, 
and one weak broad component at large values of $C$.
Therefore we divide the detected sources according to their $C$ values into three groups:
point sources with $C\leq1.32$, 
slightly extended sources with $1.32 < C\leq1.5$,
and  more extended sources with   $1.5<C\leq2.0$.
The point sources are mostly unresolved star clusters in NGC 4921 and foreground stars.
The slightly extended sources are the major population of the bright detected sources. 
They are mostly resolved GCs (with $C\leq 1.5$ and $r_{\rm eff} \leq 10$ pc) in NGC 4921. We call them compact sources hereafter. 
The $C$ values of the point and compact sources with GC-like colors show strong correlations with sharpness and  $r_{\rm eff}$.
The more extended sources are mostly blue and they are young star complexes (with $r_{\rm eff} > 10$ pc)  in NGC 4921, and we call them extended sources hereafter.

\begin{figure*}
\centering
\includegraphics[scale=1.0]{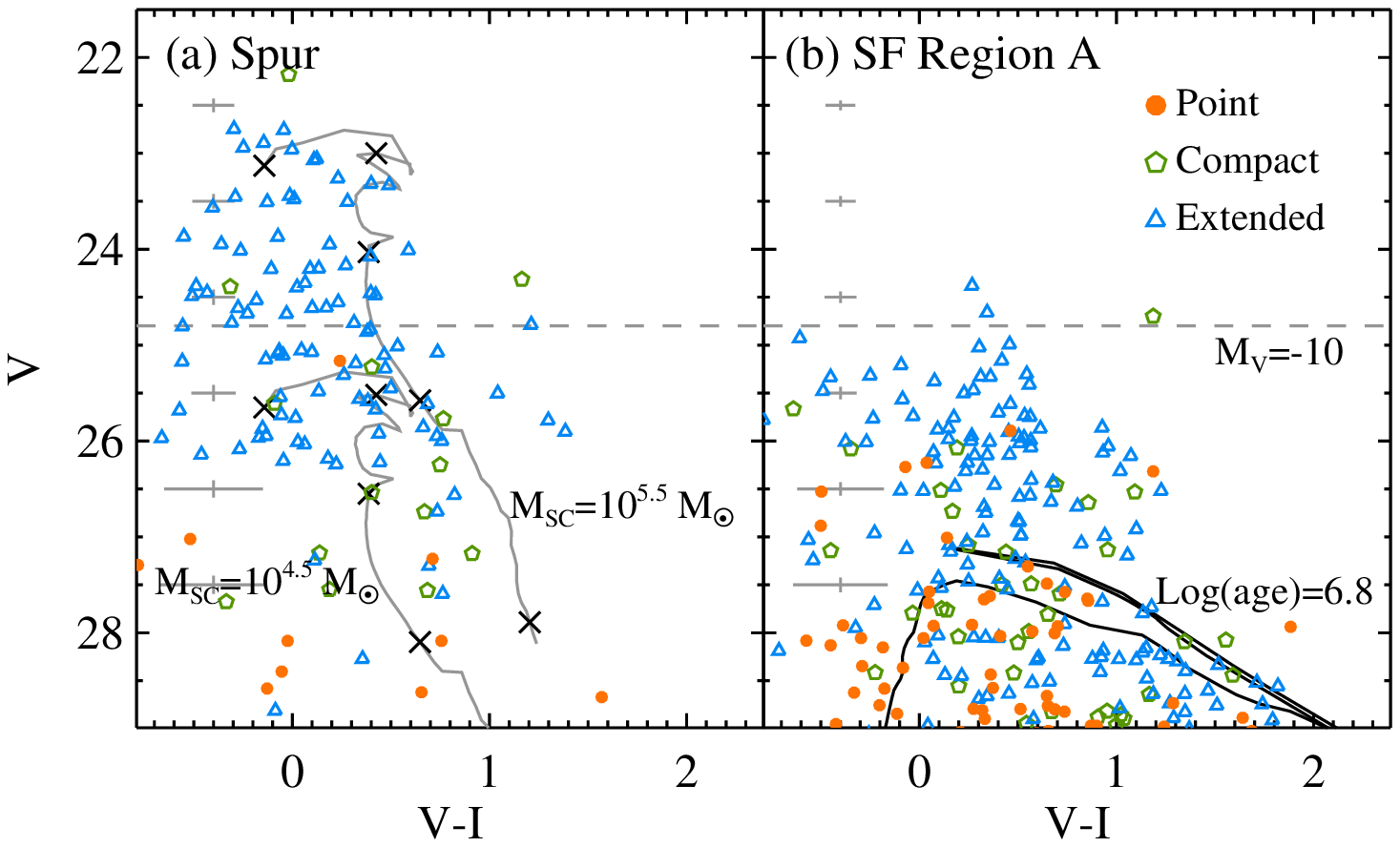} 
\caption{CMDs for the detected sources in the (a) spur region and (b) Star-forming Region A.
Solid lines represent  the SSP models for solar metallicity and masses of $10^{4.5} M_\odot$ and $10^{5.5} M_\odot$ in (a), and stellar isochrones for Log(age) = 6.8 and solar metallicity in (b),  provided by the Padova group.  Crosses along the SSP models represent the evolutionary stages for Log(age) = 6.8, 7.0, 8.0, 9.0, and 10.0 respectively. The symbols used are the same as in Fig. \ref{fig_cmd}.
Note that a majority of the detected sources in these regions are extended 
and that the brightest blue sources with $V\leq25$ mag ($M_V \leq -10$ mag) are seen mostly in the spur region. 
This indicates that massive star clusters or complexes are preferentially formed in the spur region rather than in Star-forming Region A.}
\label{fig_cmdSF}
\end{figure*}

In the case of the compact and extended sources,
we derived their total magnitude using the aperture correction method based on the relation between the aperture correction ($\Delta= mag(PSF) - mag(0\farcs5)$) and $C$ as described in \citet{lim13}.
We derived a relation using the bright isolated round sources:
$\Delta I = 1.42 \times C - 1.92$ with rms=0.1 (for $1.32<C\leq2.0$).
We adopted a constant value of 0.9 mag for the large extended sources with $C>2.0$, but the
number of these sources is very little. 
In the case of the GCs in the sample defined in this study, the value for aperture correction is  very small: a mean correction value is 0.03 mag  for $C\sim1.37$, and the maximum correction value is 0.21 mag  for $C\sim1.50$.

We performed a test to estimate the completeness of our photometry for GC candidates (compact sources), using artificial sources.
We selected $24\arcsec\times12\arcsec$ rectangular fields in the bulge ($R\leq17\arcsec$) 
and halo ($R>70\arcsec$) regions in NGC 4921.
We added 1,000 artificial compact sources with $C\sim1.37$ and $(V-I)=0.9$ and $1.1$
into each rectangular field in each image, and repeated this procedure 1000 times to generate 1000 images for each filter. 
Thus the total number of added artificial sources is 1,000,000 in each rectangular field for each filter.
We derived photometry of these sources using the same procedure as used previously.
The recovery rate (completeness), the ratio of the number of detected sources to that of the added sources, is calculated
as a function of magnitude.
  
{\bf Figure \ref{fig_compl}} shows the results of this test for the bulge and halo regions. 
The completeness limits for 50\% completeness are at $V=28.1$ mag and 29.5 mag 
($I=27.1$ mag and 28.5 mag for $(V-I)=1.0$)
for the bulge and halo regions, respectively.
The mean differences in the magnitude and color between the added sources and the detected sources are also plotted, which are smaller  than 0.1 mag for $V\leq29$ mag for the halo region.

\section{RESULTS}

\subsection{Color-Magnitude Diagrams } 

We divided the HST field into four regions for analysis:
(a) the NGC 4921 bulge region including the bar ($6\arcsec<R\leq17\arcsec$),
(b) the NGC 4921 disk region ($17\arcsec < R\leq70\arcsec$),
(c) the NGC 4921 halo region ($R>70\arcsec$ and $R_{\rm N4923}>1\arcmin$), and
(d) the NGC 4923 region ($R_{\rm N4923}\leq1\arcmin$), 
where $R$ and $R_{\rm N4923}$ are the projected angular distances in the sky from the centers of NGC 4921
and NGC 4923, respectively.
The central region at $R\leq6\arcsec$ has a high background brightness and a high incompleteness of detection so it was not used for analysis.

In {\bf Figure \ref{fig_cmd}} we display the color-magnitude diagrams (CMDs) of the detected sources with $V\leq28.0$ mag
in the four regions. We plotted the point sources, compact sources (slightly extended sources), and extended sources by filled circles, open pentagons, and open triangles, respectively.
Several notable features are found in this figure.
First, the most distinguishable feature is a narrow vertical structure reaching $V\approx 24.0$ ($M_V \approx -10.7$) mag at $(V-I)\approx 1.0$, seen
in all four regions.  This is composed mostly of compact sources. 
Their colors are similar to those of the GCs in the Milky Way.
This vertical component mainly represents the GCs in NGC 4921 and in NGC 4923, typically seen in the CMDs of the GCs in early-type galaxies.
Second, in the disk region, there are a number of blue and bright sources, reaching
$V\approx 22.7$ ($M_V \approx -12.0$) mag, more than one magnitude brighter than the brightest GCs in the same region. 
They consist mostly of extended sources. 
They are young star clusters and star complexes. The blue point sources are mostly fainter than $V\approx 26.0$ ($M_V\approx-8.7$) mag and they are probably unresolved star clusters or blue supergiants in NGC 4921.
Third, in the halo region, there are 
37 compact sources and 137 extended sources 
that are bluer than the GCs ($(V-I)<0.7$).
They are mostly  background galaxies.
%
Fourth, the GCs in NGC 4923 show a GC sequence as narrow as the one in the halo region of NGC 4921.
Sixteen extended sources bluer than the GCs in this region are most likely background galaxies.
 
\subsubsection{Star Clusters and Complexes in  the Spur Region and Star-forming Region A}

In {\bf Figure \ref{fig_cmdSF}} we plot the CMDs of the sources detected in the spur region and Star-forming Region A, as marked in Figure \ref{fig_finder}.  We have also overlayed  simple stellar population (SSP) models with solar metallicity for mass $3\times 10^4 M_\odot$ and $3\times 10^5 M_\odot$, 
provided by the Padova group \citep{gir00}.
In the case of the spur region, we plotted only the bright sources located along the dust lane with spurs, as denoted by circles in the bottom-right panel of  Figure \ref{fig_finder}. 

{\bf Figure \ref{fig_cmdSF}} shows several interesting features.
%
First, the most notable feature is that the brightest clusters in the spur region are about two magnitudes brighter than those in Star-forming Region A. 
While few sources in Star-forming Region A are brighter than $V \approx 25$ mag ($M_V \approx -9.7$ mag at the distance of NGC 4921), a number of clusters in the spur region are brighter than this magnitude, 
with some as bright as  $V \approx 23$ ($M_V \approx -11.7$) mag. 
The brightest clusters in the spur region are mostly bluer than  $(V-I)=0.5$.
About a half of the brightest clusters with $V \lesssim 25$ mag in the same region are bluer than $(V-I)=0.0$, showing
that they suffer little from interstellar reddening and that they are very young. 
The color of most bright blue star clusters in the spur region ranges from $(V-I)\approx -0.6$ to $\approx 0.4$, while those of the SSP models for the two youngest ages, Log(age)=6.8 and 7.0, are $(V-I)\approx -0.1$ and $\approx 0.4$. 
If the star clusters with $(V-I)\approx 0.5$ have an age of about 10 Myr, internal reddening must be negligible. Or, they may suffer an internal reddening of $E(V-I)\approx 0.5$, if they have an age of about 6 Myr.
Thus, most of the brightest clusters in this spur region
are thought to be younger than 10 Myr.
This indicates that massive star clusters are formed preferentially in the spur region than in Star-forming Region A, and 
that a major population of the star clusters in Star-forming Region A are older ($\approx100$ Myr) than those in the spur region. However, there are 
11 blue clusters ($(V-I)\leq0.0$) in the same region that are brighter than $V=26$ mag, showing that they were formed as recently as those in the spur region. 
Comparison with the SSP models indicates that the bright star clusters detected in the spur region have masses from $3\times 10^4 M_\odot$ to $3 \times 10^5 M_\odot$.

Second, the bright clusters (with $V\leq26$ mag) in the spur region are, on average, bluer than those in Star-forming Region A. The peak (or mean) color of the sources is
$(V-I)\approx 0.0$ for the spur region, and $(V-I)\approx 0.4$ for the star-forming region A. This indicates that the star clusters in the spur region are, on average, younger than
those in Star-forming Region A. 
Third, most of the detected sources in these regions are extended sources, so that they are star complexes or large star clusters that are larger than the GCs. 
Fourth, most of the point sources are fainter than $V\approx 27.0$ ($M_V\approx -7.8$) mag. Their location is roughly consistent with the theoretical isochrone for solar metallicity and Log(age)=6.8 from the Padova group. Thus most of the faint point sources are young massive supergiants, while some of them may be unresolved star clusters. 
These results show that massive star clusters (complexes)  were formed preferentially in the spur region compared with the normal star-forming region without spurs.

A comparison of the CMDs in {\bf Figure \ref{fig_cmdSF}} with that for the stars in   similar regions given by \citet{tik11} (their Fig. 9) show significant differences. It is noted that \citet{tik11} presented a single CMD for both regions, so that we cannot distinguish between the spur regions and star-forming regions in their Figures 7 and 8. 
Furthermore, the CMD of \citet{tik11} shows no sources brighter than $I\approx 25$ mag, while our CMDs show  a large number of brighter sources. The CMD of \citet{tik11} is similar to that of the point  and compact sources in our {\bf Figure \ref{fig_cmdSF}}. This is because \citet{tik11} selected only pointlike sources. They considered the pointlike sources in these regions as stars, and not as star clusters. 
However, most of the sources we detected in these regions are extended clusters.

\begin{deluxetable*}{lcccccccc}
\setlength{\tabcolsep}{0.05in}
\tablecaption{Summary of GMM tests$^a$ for the GC candidates in NGC 4921 and in NGC 4923}
\tablewidth{0pt}
\tablehead{ \multicolumn{3}{c}{Blue GCs} & \multicolumn{3}{c}{Red GCs}\\
Region &\colhead{Mean} &\colhead{$\sigma$} &\colhead{Number} &\colhead{Mean} &\colhead{$\sigma$} &\colhead{Number} & \colhead{$p$} &\colhead{$D$} }

\startdata
All   	& $0.937\pm0.010$ & $0.112\pm0.007$ & $493\pm30$ & $1.200\pm0.016$ & $0.112\pm0.007$ & $200\pm30$ & 7.69e-11 & $2.35\pm0.18$	\\
Bulge   	& $1.069\pm0.049$ & $0.133\pm0.014$ & $108\pm33$ & $1.209\pm0.052$ & $0.133\pm0.014$ & $55\pm33$ & 8.92e-01 & $1.05\pm0.62$ 	\\
Disk   	& $0.914\pm0.016$ & $0.113\pm0.009$ & $214\pm19$ & $1.181\pm0.026$ & $0.113\pm0.009$ & $100\pm19$& 1.89e-04 & $2.36\pm0.28$ 	\\
Halo   	& $0.898\pm0.012$ & $0.078\pm0.014$ & $126\pm4$ & $1.226\pm0.149$ & $0.078\pm0.014$ &  $7\pm4$ & 2.15e-09 & $4.18\pm2.20$ 	\\
NGC 4923  & $0.965\pm0.030$	& $0.080\pm0.021$ & $60\pm11$  & $1.141\pm0.084$ & $0.080\pm0.021$ &  $21\pm11$& 7.56e-02 & $2.21\pm1.32$	\\
\hline
\enddata
\tablenotetext{a}{Based on the homoscedastic (same variance $\sigma$) option. $p$ represents the probability for unimodality, and $D$ is the difference in color between the two peaks.}
\label{tab_cdfgmm}
\end{deluxetable*}

 \begin{figure}
\centering
\includegraphics[scale=0.9]{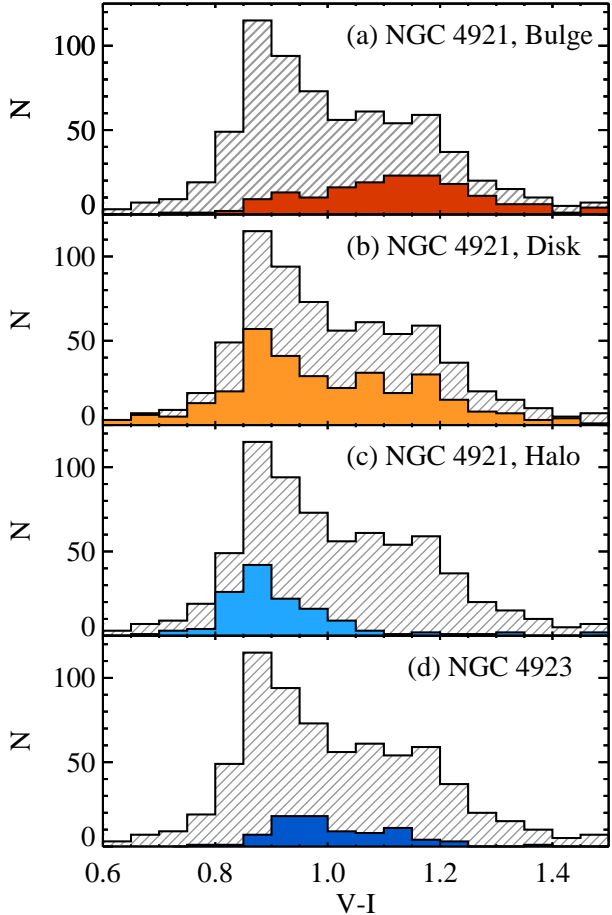} 
\caption{$(V-I)$ color histograms for the point and compact sources ($C \leq 1.5$) with $V\leq27$ mag in the (a) bulge, (b) disk, and (c) halo  of NGC 4921, and in (d) NGC 4923. In each panel, the hatched and filled histograms indicate the $V-I$ color distribution for the sources in the entire field and selected fields.
Note that most of the sources with $(0.7<(V-I)<1.4)$ are GCs.}
\label{fig_CDF}
\end{figure}

\subsection{Color Distribution of the GCs} 
In {\bf Figure \ref{fig_CDF}} we display the color distribution of the point and compact sources with $V\leq27$ mag and
$0.6<(V-I)\leq 1.5$  
 in the four regions as well as in the entire region (hatched histograms).
A majority of the sources have colors of $0.7<(V-I)<1.4$, typical for the GCs so that they are considered to be the GCs in NGC 4921.
The color distribution of the GCs in the entire region is clearly bimodal with two peaks at  $(V-I)\approx 0.9$ and $\approx 1.15$. This shows that there are two groups of GCs: blue (metal-poor) and red (metal-rich) GCs.
This figure shows significant variations among the selected regions.
First, the most dominant population in the bulge region is the red GCs
(with a peak color at $(V-I)\approx 1.15$), while  in the halo region
is the blue GCs (with a peak color at $(V-I)\approx 0.9$). 
Second, the disk GCs show  two comparable components, blue and red. 
Third, the GCs in NGC 4923 show 
a blue peak color at $(V-I)\approx 0.95$, slightly redder than the blue peak for the halo GCs in NGC 4921, $(V-I)\approx 0.9$.
NGC 4923 also shows a weaker red peak at $(V-I)\approx 1.13$.

We tested and quantified the bimodality of the color distributions using the Gaussian Mixture Modelling (GMM) program provided by \citet{mur10}, summarizing the results in Table \ref{tab_cdfgmm}.
The probability for the unimodal distribution ($p$) and the difference of two peaks ($D$) are two useful indicators  when testing the bimodality of the data. The condition, $p<0.1\%$, is used for non-unimodal distributions, and another condition, $D>2$, is for clear bimodality. We adopted a homoscedastic case (same variance) for bimodality.

These tests show that a unimodal distribution is rejected for the entire sample, the disk sample, and the halo sample of NGC 4921, with a confidence level better than 0.1\%.  The $D$ values for these samples are larger than two, showing that they show a bimodality.
On the other hand, the bulge sample shows a unimodality. The NGC 4923 sample shows $p=7.6\%$, larger than 0.1\%. Its $D$ value is $2.21\pm1.32$, indicating a bimodality, but with a large error.
The peak values and variances of the blue and red GCs are
$(V-I)=0.937\pm0.010$ and $1.200\pm0.016$ for the entire sample of NGC 4921.
Similarly,  the peak values and variances of the blue and red GCs are
$(V-I)=0.914\pm0.016$ and $1.181\pm0.026$ for the disk sample,  
$(V-I)=0.898\pm0.030$ and $1.226\pm0.149$ for the halo sample, and
$(V-I)=0.965\pm0.030$ and $1.141\pm0.084$ for the NGC 4923 sample.
The blue and red peak colors for the entire sample of NGC 4921 are $(V-I)_0 = 0.915$ and 1.178, respectively, after correction for foreground reddening $E(V-I)=0.022$. These two values correspond to the metallicities of [Fe/H] = $-1.48$ and $-0.10$, respectively, if we use the transformation for GCs,
[Fe/H]$=5.2267(V-I)_0-6.2613$ based on the Milky Way GCs \citep{wat09}.

 \begin{figure*}
\centering
\includegraphics[scale=0.9]{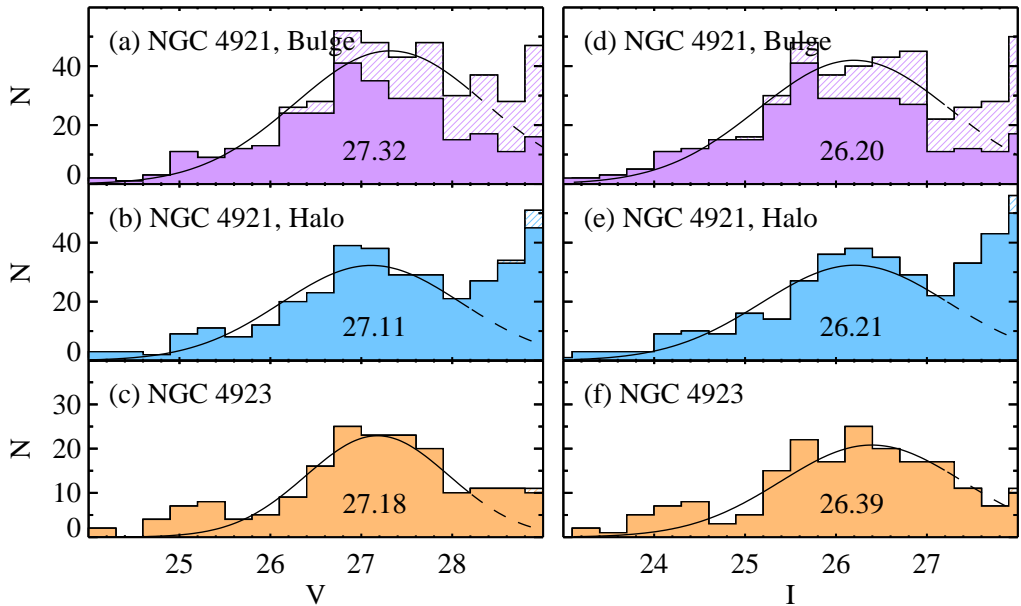} 
\caption{
$V$-band (left panels) and $I$-band (right panels) LFs of all GCs ($0.7 < (V-I)\leq1.4$) in the bulge and halo of NGC 4921, and of all GCs in NGC 4923.  
The curved lines represent the results of from Gaussian fitting. Only the bright parts of the GCLFs were used for fitting, as shown by the solid curved lines. 
Turnover magnitudes are labeled in each panel.
The 50 \% completeness limiting magnitudes are
$V=28.1$ mag ($I=27.1$ mag for $(V-I)=1.0$) for the bulge region, and $V=29.5$ mag ($I=28.5$ mag) for the halo region.
}
\label{fig_GCLFall}
\end{figure*}

\begin{deluxetable*}{lcccccc}
\setlength{\tabcolsep}{0.05in}
\tablecaption{Summary of Gaussian Fits for LFs of all GCs in NGC 4921 and in  NGC 4923}
\tablewidth{0pt}
\tablehead{ \multirow{2}{*}{Region} &  \multicolumn{3}{c}{$V$} & \multicolumn{3}{c}{$I$} \\
  & \colhead{Center} &\colhead{Width} & \colhead{$N_{\rm total}^a$} & \colhead{Center} &\colhead{Width} & \colhead{$N_{\rm total}^a$}
}
\startdata
NGC 4921 bulge, $6\arcsec < R \leq 17\arcsec$ 	& $27.32\pm0.13$ & $1.05\pm0.11$ & $395\pm37$  & $26.20\pm0.11$ & $1.06\pm0.10$ & $370\pm30$	\\
NGC 4921 halo, $70\arcsec < R \leq 160\arcsec$ & $27.11\pm0.09$ & $0.99\pm0.08$ & $267\pm20$  & $26.21\pm0.11$ & $1.03\pm0.08$ & $277\pm20$	\\
NGC 4923, $R_{N4923} \leq 1\arcmin$ 				& $27.18\pm0.09$ & $0.79\pm0.09$ & $151\pm15$  & $26.39\pm0.13$ & $1.00\pm0.12$ & $174\pm17$ 	\\

\hline
\enddata
\tablenotetext{a}{The errors for the numbers are from the fitting errors for the integration of Gaussian functions provided by IDL/mpfitexpr.}
\label{tab_gclfN4921all}
\end{deluxetable*}

\subsection{GCLFs} 

 \begin{figure*}
\centering
\includegraphics[scale=0.9]{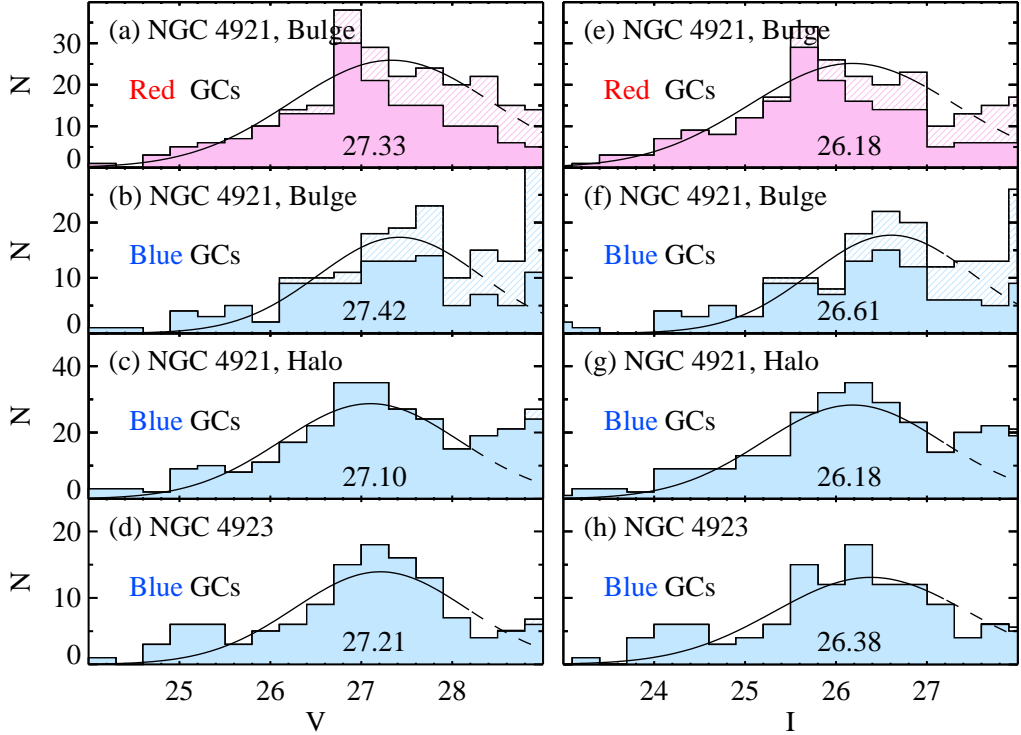} 
\caption{$V$-band (left panels) and $I$-band (right panels) LFs of the red GCs 
 in the bulge of NGC 4921, the blue GCs 
 in the bulge and in the halo of NGC 4921, and  the blue GCs in NGC 4923.  
The filled and hatched histograms represent the GCLFs before and after completeness correction, respectively.
The curved lines represent the results from Gaussian fitting. Only the bright parts of the GCLFs were used for fitting, as shown by the solid curved lines. 
Turnover magnitudes are labeled in each panel.
}
\label{fig_GCLFblue}
\end{figure*}


We selected the point sources and compact sources 
with $0.7<(V-I)\leq1.4$ to be GCs.
We then  derived $V$ and $I$-band GCLFs for the NGC 4921 bulge region,  
the NGC 4921 halo region (excluding NGC 4923), and  NGC 4923. 
{\bf Figure \ref{fig_GCLFall}} displays the LFs for all GCs in each region. We plotted the GCLFs before and after completeness correction 
using the completeness test results. 
The completeness correction is significant in the faint end for the bulge region, but is negligible for the halo region.
The GCLFs show a turnover (peak) at $V \approx 27.2$ mag ($I \approx 26.2$ mag), 
and decrease thereafter. 
The excess in the faint sources with $V>28.2$ mag ($I>27.2$ mag) is most likely due to background sources.
We fit the GCLFs for all GCs with $V\leq28$ mag ($I\leq27$ mag) where incompleteness of our photometry is not significant, and listed the fitting parameters in {\bf Table \ref{tab_gclfN4921all}}:
$V{\rm (max)} = 27.32\pm0.13$ mag and $\sigma = 1.05\pm0.11$ for the bulge sample,
$V{\rm (max)} = 27.11\pm0.09$ mag and $\sigma = 0.99\pm0.08$ for the halo sample, and
$V{\rm (max)} = 27.18\pm0.09$ mag and $\sigma = 0.79\pm0.09$ for the NGC 4923 sample.
Similarly for $I$-band data, we obtain
$I{\rm (max)} = 26.20\pm0.11$ mag and $\sigma = 1.06\pm0.10$ for the bulge sample,
$I{\rm (max)} = 26.21\pm0.11$ mag and $\sigma = 1.03\pm0.08$ for the halo sample, and
$I{\rm (max)} = 26.39\pm0.13$ mag and $\sigma = 1.00\pm0.12$ for the NGC 4923 sample.
Thus the turnover magnitudes for all GCs in the different regions are similar.

\begin{deluxetable*}{lcccccccc}
\setlength{\tabcolsep}{0.05in}
\tablecaption{Summary of Gaussian Fits for LFs of the Blue and Red GCs in NGC 4921 and in NGC 4923}
\tablewidth{0pt}
\tablehead{ \multirow{2}{*}{Galaxy} & \multirow{2}{*}{Color} & \multirow{2}{*}{Region} &  \multicolumn{3}{c}{$V$} & \multicolumn{3}{c}{$I$} \\
  & & & \colhead{Center} &\colhead{Width} & \colhead{$N_{\rm total}$} & \colhead{Center} &\colhead{Width} & \colhead{$N_{\rm total}$}
}
\startdata
NGC 4921 &red, $1.05<V-I\leq1.40$& bulge, $6\arcsec < R \leq 17\arcsec$ 	
& $27.33\pm0.19$ & $1.11\pm0.16$ & $240\pm32$  
& $26.18\pm0.22$ & $1.13\pm0.17$ & $237\pm35$	\\
\hline

NGC 4921 &blue, $0.70<V-I\leq1.05$& bulge, $6\arcsec < R \leq 17\arcsec$ 	
& $27.42\pm0.25$ & $0.89\pm0.21$ & $129\pm14$ 
& $26.61\pm0.27$ & $0.89\pm0.27$ & $132\pm15$	\\
 
NGC 4921 &blue, $0.70<V-I\leq1.05$& halo, $70\arcsec < R \leq 160\arcsec$ 
& $27.10\pm0.11$ & $0.99\pm0.12$ & $236\pm22$  
& $26.18\pm0.11$ & $0.99\pm0.12$ & $234\pm22$	\\

NGC 4923 &blue, $0.70<V-I\leq1.05$&$R_{N4923} \leq 1\arcmin$ 				
& $27.21\pm0.16$ & $0.95\pm0.17$ & $110\pm15$  
& $26.38\pm0.24$ & $1.06\pm0.23$ & $116\pm21$ 	\\

\multicolumn{3}{l}{Mean of NGC 4921 halo and NGC 4923 for blue GCs}				
& $27.14\pm0.09$ & $0.98\pm0.10$ &   
& $26.21\pm0.10$ & $1.00\pm0.11$ & 	\\
\hline
\enddata
\label{tab_gclfN4921blue}
\end{deluxetable*}

 \begin{figure*}
\centering
\includegraphics[scale=0.9]{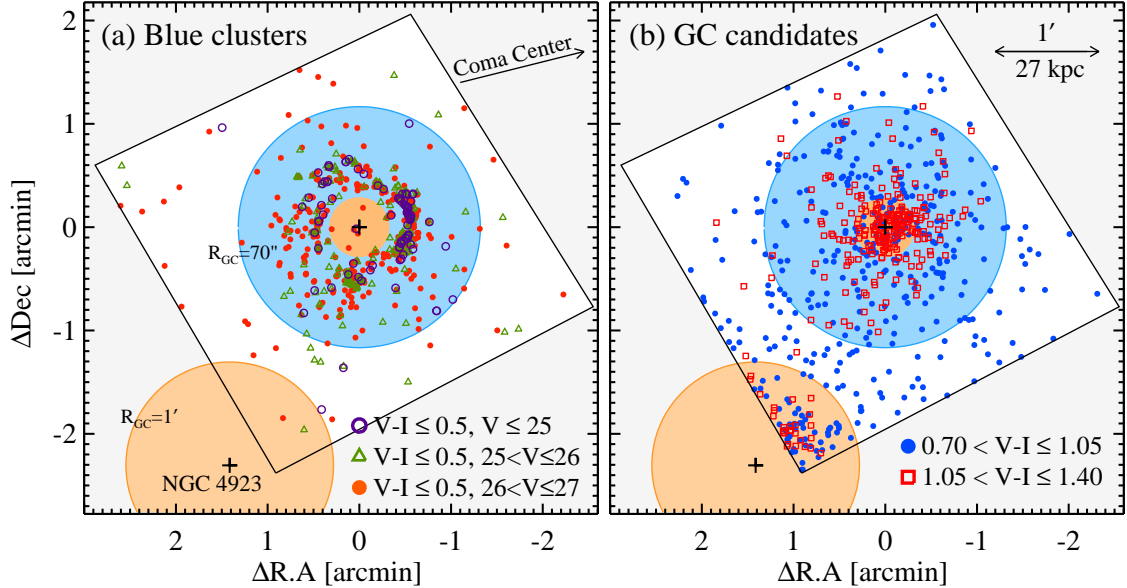} 
\caption{(Left panel) Spatial distribution of very blue sources ($(V-I)\leq 0.5$) for $V<\leq 25$ mag (open circles),
$25<V\leq 26$ mag (open triangles), and $26<V\leq27$ mag (filled circles) in NGC 4921 and NGC 4923. 
(Right panel) Spatial distribution of bright GCs  with $V\leq27$ mag. 
Filled circles and open squares represent blue 
GCs ($0.70<(V-I)\leq1.05$) and red GCs ($1.05<(V-I)\leq1.40$), respectively.
}
\label{fig_spat}
\end{figure*}

 \begin{figure}
\centering
\includegraphics[scale=0.9]{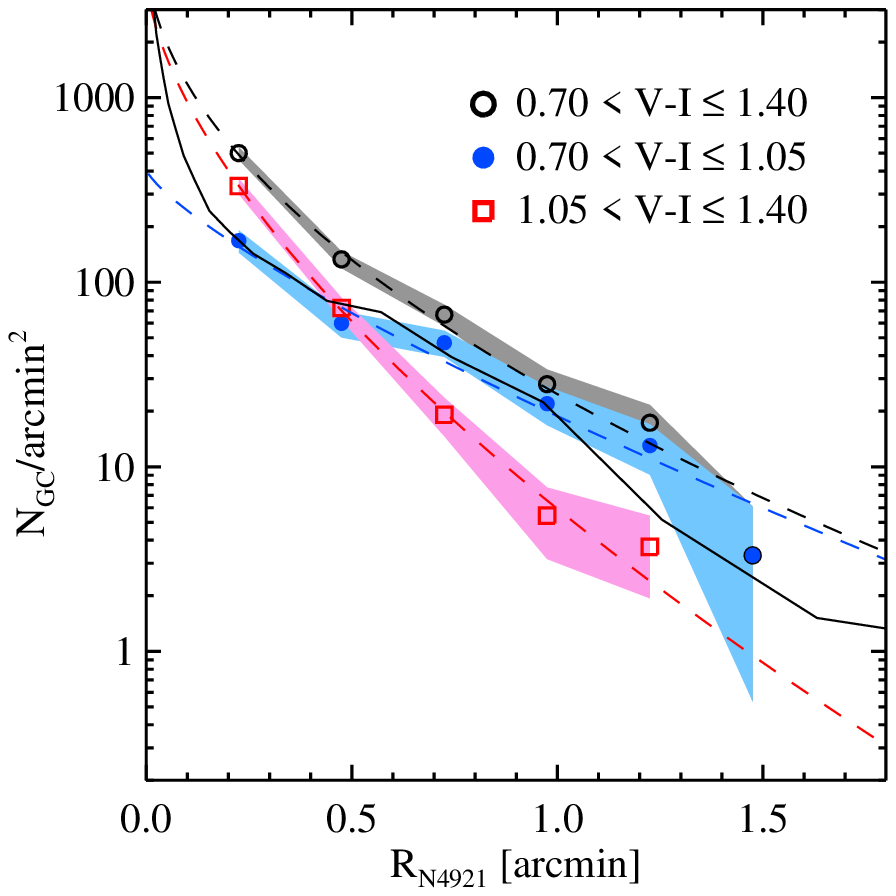} 
\caption{Radial number density distributions of the bright GCs 
(with $V\leq27$ mag) in NGC 4921. 
Open circles, filled circles, and open squares represent the all GCs, blue 
GCs, 
and red GCs, 
respectively.
We subtracted the contribution due to the background  using the control field at $R > 1\farcm7$.
Note 
that the red GCs are located mostly at $R<1'$, while the distribution of the blue GCs is much more extended.
The solid line is the $V$-band surface brightness profile shifted arbitrarily.
The dashed lines represent the Sersic law fitting results for the GCs: $n=1.93\pm0.62$ for all GCs, $n=1.28\pm0.52$ for the blue GCs, and $n=1.60\pm0.65$ for the red GCs. 
\label{fig_radialden}}
\end{figure}

We divided the GC sample into two subgroups according to their color: 
blue (metal-poor) GCs are those with $0.7<(V-I)\leq 1.05$, and  red (metal-rich) GCs are those with $1.05<(V-I)\leq1.40$. 
We derived the GCLFs for the blue GCs 
in the NGC 4921 bulge and halo regions, and in NGC 4923.
In the case of the NGC 4921 bulge region, we also presented the GCLF for the red GCs. 
Fitting results for these GCLFs with Gaussian functions are summarized  in Table \ref{tab_gclfN4921blue}.
%
{\bf Figure \ref{fig_GCLFblue}} shows $V$ and $I$-band GCLFs for the blue and red GCs in these regions.
In the NGC 4921 halo region and in NGC 4923, the turnover magnitudes for the blue GCs
are similar to those of all GCs. This is because these regions are dominated by the blue GCs.

The blue GCs in the NGC 4921 halo region and in NGC 4923 are considered to be the best
sample for estimating the turnover magnitude of the GCLFs for the blue GCs.
In the case of disk galaxies, GCs are mainly composed of metal-poor GCs in the halo and of metal-rich GCs in the bulge. The turnover magnitudes of the metal-rich GCs are known to be fainter than those of the metal-poor GCs in some galaxies \citep{dic06,rej12}. 
Therefore the GCLFs of the metal-poor GCs are considered a better standard candle than those of the entire GCs or the GCLFs of the metal-rich GCs \citep{dic06,rej12}.

The turnover magnitudes of the GCLFs for the blue GCs
are determined to be
$V{\rm (max)} = 27.10\pm0.11$ mag and $I{\rm (max)} = 26.18\pm0.11$ mag for the NGC 4921 halo region, and 
$V{\rm (max)} = 27.21\pm0.16$ mag and $I{\rm (max)} = 26.38\pm0.24$ mag for NGC 4923. 
The mean values of these two are
$V{\rm (max)} = 27.14\pm0.09$ mag and $I{\rm (max)} = 26.21\pm0.10$ mag. 
In contrast, the red GCs in the NGC 4921 bulge region are considered to be the best
sample for estimating the turnover magnitude of the GCLFs for the red GCs:
 $V{\rm (max)} = 27.33\pm0.19$ mag and $I{\rm (max)} = 26.18\pm0.22$ mag. 
Note that the turnover magnitudes for the red GCs in the NGC 4921 bulge are not much different from those for the blue GCs in NGC 4921. This is in strong contrast to the GCs in the Milky Way.

\subsection{Spatial Distribution of the GCs} 
In {\bf Figure \ref{fig_spat}(b)} we plot the spatial distribution of 
the bright GCs with $V\leq27$ mag in NGC 4921 and  in NGC 4923. The blue GCs and red GCs are marked with different symbols. 
{  We also displayed  the spatial distribution of the very blue compact and extended clusters with $V\leq28$ mag and $(V-I)<0.5$ (bluer than the GCs in the same field) for comparison in {\bf Figure \ref{fig_spat}(a)}.
Note that 
the red GCs of NGC 4921 are found mostly in the bulge and disk region, showing a strong central concentration,
while the blue GCs of NGC 4921 show a much weaker central concentration. 
NGC 4923 contains both blue and red GCs.
The spatial distributions of the GCs are significantly different from those of the very blue sources.
The very blue sources show strong concentrations
in the spiral arm regions, and only few of them are found in the bulge region.  This suggests that most of these sources are young star clusters in NGC 4921. 
It is noted that the concentration of these blue sources is significantly stronger  in the spur region than  in other regions.
On the other hand, the distribution of these blue sources is almost uniform in the halo region and they have color $(V-I)>-0.1$ (see Figure 4), indicating that the very blue sources in the halo region are mainly background sources.
}


Using the same samples as used for studying the GCLFs, we derived the radial number density distribution for
the blue GCs, red GCs, and all GCs, respectively, 
as shown in {\bf Figure \ref{fig_radialden}}.
We used only the bright GCs with $V\leq27$ mag.
 We estimated the background levels using the sources with the same ranges of colors and magnitudes at $R>1\farcm7$ excluding the $1\arcmin$ region from the center of NGC 4923. 

The radial profile of the red GCs in {\bf Figure \ref{fig_radialden}} is steep in the inner region ($R\leq0\farcm95$), flattens in the outer region at $0\farcm95<R\leq1\farcm15$, and declines rapidly at $R>1\farcm15$.
The radial profile of the blue GCs is much flatter than that of the red GCs. 
These results show that the red GCs in the inner region  mostly belong to the bulge of NGC 4921, while most of the red GCs at $R>0\farcm8$ are most likely disk clusters.
The blue GCs in the outer region mostly belong to the halo.

The radial profiles can be roughly described  by the Sersic law \citep{ser63}. We fit these profiles with a Sersic law with an index $n$ \citep{ser63}, plotting them by the dashed lines:
$n=1.93\pm0.62$ and $r_{\rm eff}=0\arcmin.41\pm0\arcmin.04$ for all GCs, $n=1.28\pm0.52$ and $r_{\rm eff}=0\arcmin.66\pm0\arcmin.10$ for the blue GCs, 
and $n=1.60\pm0.65$ and $r_{\rm eff}=0\arcmin.26\pm0.\arcmin.05$ for the red GCs.
Thus the effective radius of the blue GC system is more than twice larger than that of the red GC system.

 \begin{figure}
\centering
\includegraphics[scale=0.9]{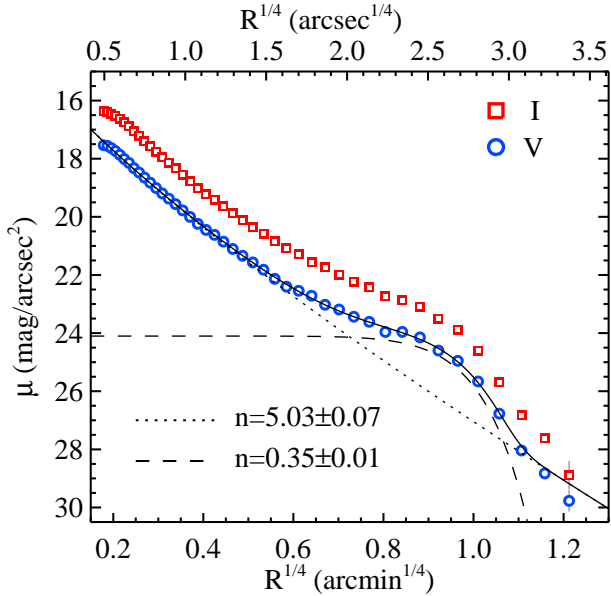} 
\caption{Radial surface brightness profiles of NGC 4921 for $V$ (circles) and $I$ bands (squares).
Two component Sersic profile fits for $V$-band 
at $0.3 <(R[{\rm arcmin}])^{1/4}  \leq 1.2$ range are shown by the dotted (spheroidal component), dashed (disk component), and solid lines (sum).
\label{fig_surf}}
\end{figure}

\subsection{Surface Photometry of NGC 4921}

We derived radial profiles of $V$ and $I$-band surface brightness of NGC 4921 using  
IRAF/ELLIPSE from the same HST images. We estimated the background level using the outermost regions in the image ($R>1\farcm7$), and masked out large sources except for NGC 4921 before applying ELLIPSE. 
{ We fixed the values of the galaxy center and set other parameters to be free during the ELLIPSE fitting.}

{\bf Figure \ref{fig_surf}} shows the $V$ and $I$-band surface brightness profiles of NGC 4921. 
Both profiles look similar.
The surface brightness profiles show a break at
$R^{1/4} \approx 0.6$ ($R\approx 8\arcsec$).
These profiles become flatter and show another break
at $R^{1/4} \approx 1.0$ ($R\approx 60\arcsec$),
becoming very steep at $R>1\arcmin$. 
The inner part of the radial profiles represents a bulge component, and the outer part shows a disk component. 
We fit the surface brightness profiles with a  two component Sersic law \citep{ser63}. The bulge is fit well using a deVaucouleurs law ($n=5.45\pm0.07$) and the outer region is fit using an  exponential disk law ($n=0.34\pm0.01$).
The $V$-band disk surface brightness reaches
$\mu_V =30 $ mag arcsec$^{-2}$ at $R\approx 2\arcmin$, which is close to the boundary of the field. 
We plotted the  $V$-band surface brightness profile of
NGC 4921 for comparison with the radial number density profiles of the  GCs in {\bf Figure \ref{fig_radialden}}.
The  $V$-band surface brightness profile of the inner
 region ($R<0\farcm8$) of NGC 4921 is similar to the radial number density profile of the blue GCs in the same galaxy, while the surface brightness profile of the outer region ($0\farcm5 <R< 1\farcm0$) is steeper than that of the blue GCs.

\section{DISCUSSION}

\subsection{Distances to NGC 4921 and NGC 4923}


The distance to NGC 4921 is not yet well-known.
Using the GCLFs derived from the HST images,
\citet{tik11}
estimated a distance to the pair of NGC 4921 and NGC 4923. 
They found turnover magnitudes of $V{\rm (max)} =27.4$ mag for the inner region of NGC 4921 ($10''<R<25''$), and $V{\rm (max)} =27.5$ mag for NGC 4923. They did not provide any errors for these values.
Adopting $M_V {\rm (max) } = -7.5$ mag, they obtained a distance modulus of 
$(m-M)_0 =34.9$ for NGC 4921 and $(m-M)_0 =35.0$ for NGC 4923. 
They then presented a mean value of $97\pm5$ Mpc for the distance. 
However, note that they recorded a value of  $V{\rm (max)} =27.5$ mag for the peak magnitude of NGC 4923 in their manuscript, but Figure 4 in their paper shows that the peak they marked for NGC 4923 is at $V{\rm (max)} =27.0$ mag, 0.5 mag brighter than the value they used for analysis. 
If the latter is adopted, the distance modulus for NGC 4923 will be $(m-M)_0 =34.5$, and the mean distance modulus of the two galaxies will be $(m-M)_0 =34.65$. 
{  If the value for NGC 4923 in their Figure 4 is adopted, the values of the turnover magnitudes for all the GCs in the bulge region  ($V{\rm (max)} =27.32\pm0.13$ mag), and in NGC 4923 ($V{\rm (max)} =27.18\pm0.09$ mag)  in this study
are in a reasonable agreement with those in \citet{tik11}}.

\begin{figure}
\centering
\includegraphics[scale=0.8]{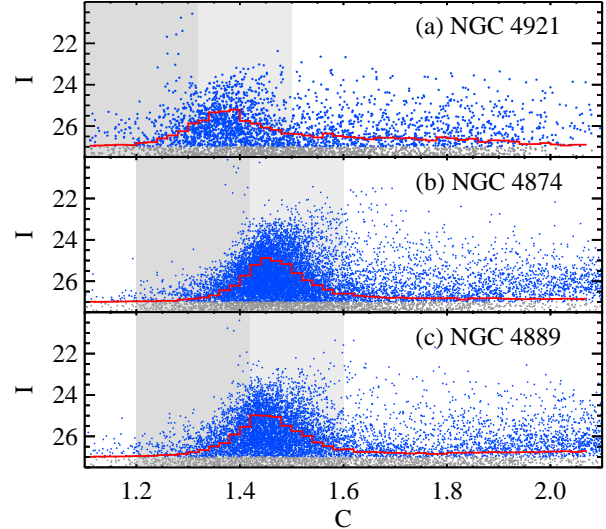} 
\caption{$I$-band magnitude vs. Concentration index ($C$) diagram and histograms for the detected sources in NGC 4921 (a), NGC 4874 (b), and NGC 4889 (c). 
The gray region and  light gray region represent the point sources and the compact sources, respectively. Note that each galaxy shows a dominant concentration at $C \approx 1.38$ (NGC 4921), and
$\approx 1.46$ (NGC 4874 and NGC 4889). The sources with $C<1.6$ in NGC 4874 and NGC 4889 are mostly resolved GCs.
}
\label{fig_CI3}
\end{figure}



We adopt the calibration of the turnover $V$-band magnitudes for the metal-poor (blue) GCs given by \citet{dic06} and \citet{rej12}:
$M_V {\rm (max)} = -7.66\pm0.09$ mag. 
This leads to the $I$-band calibration of $M_I {\rm (max)} = -8.56\pm0.09$ mag for $(V-I)_0=0.9$ (the peak color of the blue GCs in NGC 4921).
Applying the $V$-band calibration for the blue GCs, 
we determine the distances to NGC 4921 and NGC 4923: 
$(m-M)_0 =34.73\pm0.14$ ($d=88.3\pm5.8$ Mpc), and
$(m-M)_0 =34.84\pm0.18$ ($d=93.0\pm7.9$ Mpc), respectively.
If we use the $I$-band calibration, we obtain similar values:
$(m-M)_0=34.73\pm0.14$ ($d=88.3\pm5.8$ Mpc) for NGC 4921, and
$(m-M)_0=34.93\pm0.26$  ($d= 96.8\pm11.4$ Mpc) for NGC 4923.


\subsection{Comparison with Coma cD Galaxies and the Hubble Constant}

\begin{figure*}
\centering
\includegraphics[scale=1.0]{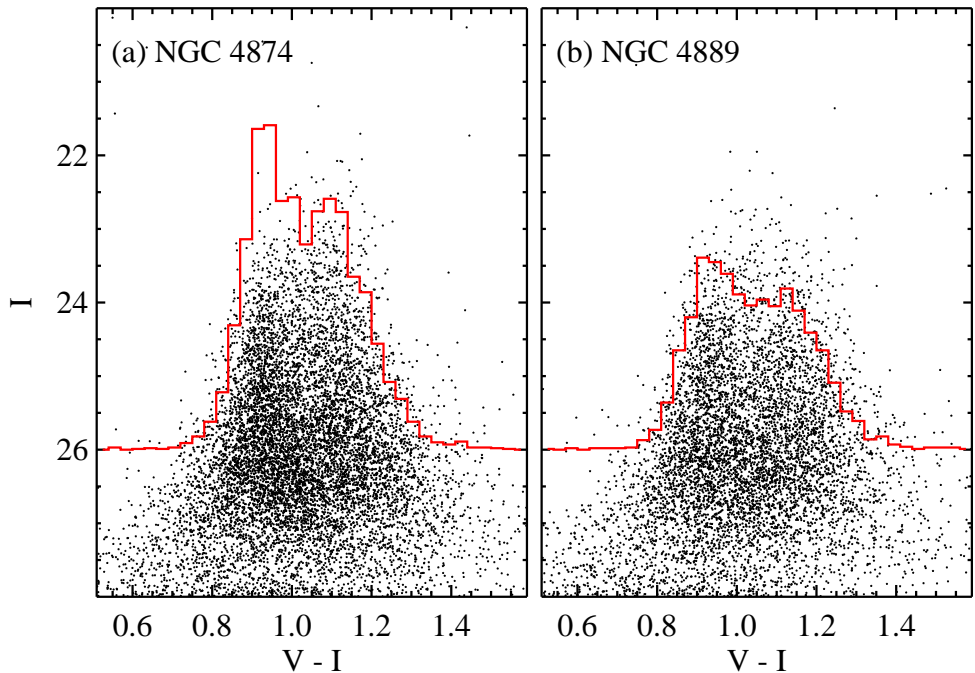} 
\caption{$I-(V-I)$ CMDs for the point and compact sources (slightly extended sources, $C \leq 1.6$) in the $0\farcm1 < R \leq 2\farcm0$ region of NGC 4874 (a) and NGC 4889 (b). 
The solid lines represent the color distribution of the bright sources with $I\leq26.0$.
Note that most of the sources with $0.7<(V-I)<1.4$ are GCs.}
\label{fig_cmdcoma}
\end{figure*}


Coma has long been used as an excellent target to estimate the Hubble constant. However it is about five times more distant than Virgo, so the determination of the distances to its member galaxies is much more difficult. The GCLF is one of the useful distance indicators for Coma galaxies. Deep HST/ACS images enabled us to measure the turnover magnitudes of the GCLFs for NGC 4921 and NGC 4923 much more accurately than for other Coma galaxies studied previously \citep{kav00,har09}.
For comparison, we studied two more Coma galaxies for which deep images are available in the archive: NGC 4874 and NGC 4889, two cD galaxies in the central region of Coma.

We analyzed the HST/ACS F475W$(g)$ and F814W$(I)$ images of  NGC 4874 and NGC 4889, in the archive (PIDs : 11711 and 10861).
We combined the images of these galaxies 
with a pixfrac value of 1.0, and a pixel scale of $0\farcs05$ for F475W and $0\farcs03$ for  F814W. 
The total exposure times of NGC 4874 are 5,071 s for F475W and 11,825 s for  F814W,  and 
those of NGC 4889 are 4,770 s for F475W and 9,960 s for  F814W. These exposure times are  much shorter than those for the NGC 4921 images. However, the images of these elliptical galaxies suffer much less crowding than those of  NGC 4921, which is an almost face-on spiral galaxy. 
We obtained photometry of the sources in these images using the same procedure as for NGC 4921. We converted F475W/F814W systems onto the Johnson-Cousins $VI$ system using the transformation relations given in \citet{ber09}.

\begin{figure*}
\centering
\includegraphics[scale=0.9]{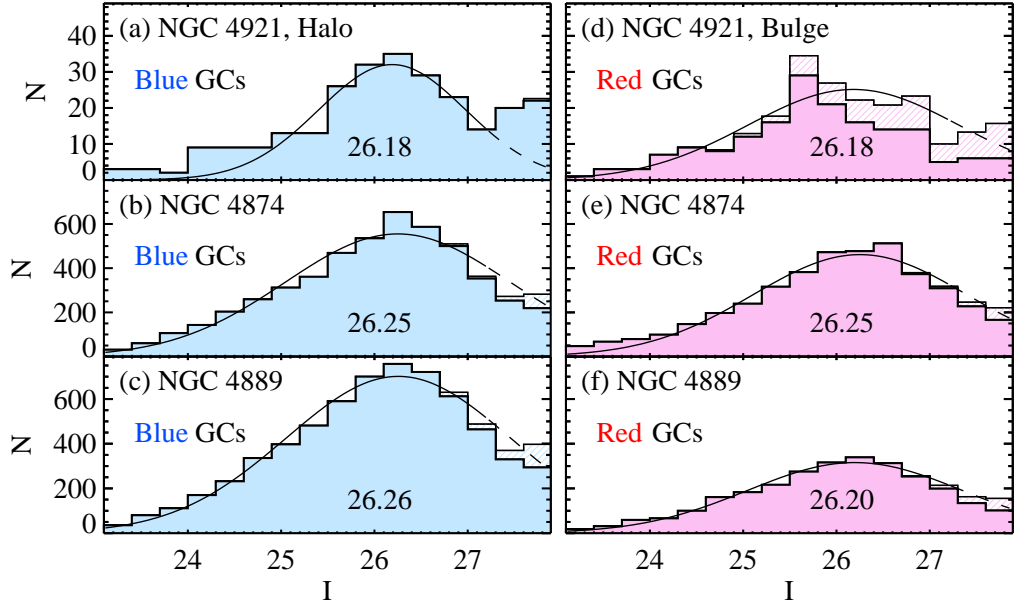} 
\caption{(Left panels) $I$-band LFs of the blue 
GCs   
 in the halo of NGC 4921, 
 and in the  regions at $0\farcm1 \leq R < 2\farcm0$ of NGC 4874 and NGC 4889. 
(Right panels) $I$-band LFs of  the red GCs  
in the bulge of NGC 4921, and in the  regions at $0\farcm1 \leq R < 2\farcm0$ of NGC 4874 and NGC 4889. 
The filled and hatched  histograms represent the GCLFs before and after completeness correction, respectively.
The curved lines represent the Gaussian fitting results. 
Only the bright parts of the GCLFs were used for fitting, as shown by the solid curved lines. 
The 50\% completeness limiting magnitude for NGC 4874 and NGC 4889 is $I\approx28.0$ mag.
}
\label{fig_GCLF3bluered}
\end{figure*}

\begin{deluxetable*}{lccccc}
\tabletypesize{\footnotesize}
\setlength{\tabcolsep}{0.05in}
\tablecaption{Summary of Gaussian Fits for GCLFs of Coma Galaxies}
\tablewidth{0pt}
\tablehead{ \multirow{2}{*}{Galaxy} & \multirow{2}{*}{Color} & \multirow{2}{*}{Region} &   \multicolumn{3}{c}{$I$} \\
&&& \colhead{Center} &\colhead{Width} & \colhead{$N_{\rm total}$}
}
\startdata
\multicolumn{6}{l}{Blue GCs}\\
\hline
NGC 4921 &$0.70<V-I\leq1.05$&$70\arcsec < R \leq 160\arcsec$	
& $26.18\pm0.11$ & $0.99\pm0.12$ & $234\pm22$	\\
NGC 4923 &$0.70<V-I\leq1.05$&$R \leq 1\arcmin$ 		
& $26.38\pm0.24$ & $1.06\pm0.23$ & $116\pm21$ 	\\
NGC 4874 &$0.70<V-I\leq1.05$& $6\arcsec < R \leq 120\arcsec$
& $26.25\pm0.04$ & $1.21\pm0.04$ & $5595\pm148$	\\

NGC 4889 &$0.70<V-I\leq1.05$& $6\arcsec < R \leq 120\arcsec$	
& $26.26\pm0.03$ & $1.20\pm0.03$ & $7047\pm167$	\\
\hline
\multicolumn{3}{l}{Weighted mean(3) excluding NGC 4923} 	& $26.25\pm0.03$ & $1.19\pm0.02$ & 	\\
\multicolumn{3}{l}{Weighted mean(4) including NGC 4923}   	&  $26.25\pm0.03$ & $1.21\pm0.02$ & 	\\
\multicolumn{3}{l}{For $(V-I)_0=0.9$ and $M_I=-8.56\pm0.09$}   	& 
\multicolumn{3}{l}{$(m-M)_0 = 34.80\pm0.09$} 	\\
\multicolumn{3}{l}{For $M_V=-7.66\pm0.09$}   	& \multicolumn{3}{l} {$(m-M)_0 = 34.78\pm0.09$} \\
\hline\hline
\multicolumn{6}{l}{Red GCs}\\
\hline
NGC 4921 &$1.05<V-I\leq1.40$&$6\arcsec < R \leq 17\arcsec$	
& $26.18\pm0.22$ & $1.13\pm0.17$ & $237\pm35$	\\

NGC 4874 &$1.05<V-I\leq1.40$& $6\arcsec < R \leq 120\arcsec$	
& $26.25\pm0.04$ & $1.15\pm0.04$ & $4420\pm121$	\\

NGC 4889 &$1.05<V-I\leq1.40$& $6\arcsec < R \leq 120\arcsec$	
& $26.20\pm0.05$ & $1.15\pm0.05$ & $3031\pm97$	\\
\hline
\multicolumn{3}{l}{Weighted mean(3)}     	& $26.23\pm0.03$ & $1.15\pm0.03$ & 	\\
\enddata
\label{tab_gclfComa3}
\end{deluxetable*}

{\bf Figure \ref{fig_CI3}} plots  $I$-band magnitudes versus concentration index $C$ values for the sources with $I \leq 27.0$ mag in NGC 4874 and NGC 4889 in comparison with NGC 4921. The histograms of the $C$ values are also overlaid.
The histograms of the $C$ values for NGC 4874 and NGC 4889 show a vertical sequence of bright sources with $I<24$ mag at $C\approx 1.40$ (the gray shaded regions). The sources in this sequence are point sources, most of which are foreground stars.
This value is about 0.2 larger than that for NGC 4921.
This difference is considered to result from the difference in the value of pixfrac used for the image combining (0.7 for NGC 4921, and 1.0 for NGC 4874 and NGC 4889).
The histograms of the $C$ values 
for NGC 4874 and NGC 4889 show a dominant peak at
$C \approx 1.45$, which is about 0.05 larger than the value for the point sources in the same galaxies.
The sources in this dominant component are slightly extended (compact) and those with $C<1.6$ are mostly GCs in these galaxies. Thus a significant fraction of the GCs in these galaxies are resolved in the images we used. 
We used $I$-band magnitudes for the analysis of the GCLFs of these galaxies in the following,  because the $I$-band data for these galaxies were
obtained using the same F814W filters as used for NGC 4921 and NGC 4923.

{\bf Figure \ref{fig_cmdcoma}} displays the CMDs of the detected sources with $1.2<C\leq1.6$ in NGC 4874 and in NGC 4889. These sources are located at $0\farcm1<R<2\farcm0$
from the center of each galaxy. 
The CMDs show that most of the bright sources ($I\leq27.0$ mag) have colors
$0.7<(V-I)<1.4$ and that they are mostly GCs in these galaxies. 
We also plotted the color distribution of the bright sources with $I\leq26$ mag with histograms.
The color distribution is clearly bimodal, showing peaks at
$(V-I)\approx 0.95$ and 1.1 for NGC 4874 and NGC 4889.
We divided the GC sample into two
groups according to their color: 
blue GCs ($0.70<(V-I)\leq1.05$) and red GCs $1.05<(V-I)\leq1.4$.
Then we derived the GCLFs for these blue and red GCs in NGC 4874 and NGC 4889.
%

In {\bf Figure \ref{fig_GCLF3bluered}} 
we plot  the $I$-band GCLFs for the blue GCs and red GCs in NGC 4874 and NGC 4889.
The GCLFs for the blue GCs in the NGC 4921 halo and those for the red GCs in the NGC 4921 bulge are also included for comparison. 
We plotted the GCLFs before and after completeness correction by filled and hatched histograms, respectively.
In the case of NGC 4874 and NGC 4889, we used the sources at $0\farcm1< R\leq 2\farcm0$ in deriving their GCLFs.
The 50\% completeness limit of these sources is $I\approx 28.0$ mag for NGC 4874 and 4889.
The GCLFs follow approximately Gaussian functions.
We fitted the GCLFs with Gaussian functions,  listing the results in {\bf Table \ref{tab_gclfComa3}}. We used only the bright part of the GCLFs with $I\leq27.2$ for fitting, where incompleteness is not significant.

A few important features in these figure and table are noted.
First, the GCLFs of NGC 4874 and NGC 4889 show a clear peak at $I \approx 26.3$ mag and show a Gaussian form more than one magnitude thereafter. 
Second, the  $I$-band  turnover magnitudes for the blue GCs in NGC 4921 and in NGC 4923 are remarkably similar to those for the two cD galaxies:
$I{\rm (max)} = 26.18\pm0.11$ mag for NGC 4921,  
$26.38\pm0.24$ mag for NGC 4923,
$26.25\pm0.04$ mag for NGC 4874 and  
$26.26\pm0.03$ mag for NGC 4889. 
We derive a weighted mean value for these four galaxies, 
$I$ ({\rm max}) $ = 26.25\pm0.03$ mag.
Here the error is a mean value of the weighted errors for the four measurements.
Third, the turnover magnitudes for the blue GCs are very similar to those for the red GCs in NGC 4874 and NGC 4889: $I {\rm (max)}  = 26.25\pm0.04$ mag for NGC 4874 and  
 $26.20\pm0.05$ mag for NGC 4889.
Fourth, the turnover magnitudes for the red GCs in NGC 4874 and NGC 4889 are also similar to those for the red GCs in the NGC 4921 bulge, $I$ ({\rm max}) $ = 26.18\pm0.22$ mag.  The weighted mean value for the red GCs in the NGC 4921 bulge, NGC 4874 and NGC 4889 is  $I$ ({\rm max})  $= 26.23\pm0.03$ mag.
Thus the turnover magnitudes for the blue GCs are little different from those for the red GCs in these galaxies.
 Fifth, the weighted mean values of the Gaussian widths for these galaxies are
$\sigma=1.21\pm0.02$ for the blue GCs, and
$\sigma=1.15\pm0.03$ for the red GCs.
These values are consistent with those in the previous studies for the Milky Way Galaxy and M87 \citep{pen09,rej12}. However, they are smaller than the value for Coma galaxies, $\sigma \approx 1.4\pm0.1$,  \citet{kav00}, \citet{woo00}, and \citet{har09} derived from shallower data.

\subsubsection{GCLFs for Virgo and Coma Galaxies}

The best data for the GCLFs of massive early-type galaxies in the Virgo cluster is available for M87, a cD galaxy located close to the cluster center.
\citet{pen09} studied the GCLF for M87 from HST/ACS  F606W  and  F814W  images, presenting turnover magnitudes:
$I_0 ({\rm max}) =22.53\pm0.05$ mag and $\sigma =1.37\pm0.04$ 
for the entire sample of GCs,
$I_0 ({\rm max})  =22.24\pm0.06$ mag and $\sigma =1.25\pm0.05$ 
for the blue GCs ($(V-I)<1.04$), and
$I_0 ({\rm max}) =22.77\pm0.09$ mag and $\sigma =1.45\pm0.06$ 
for the red GCs ($(V-I)>1.04$).
The result for all GCs  is consistent with that given by \citet{kun99}, $I_0 ({\rm max})  =22.55\pm0.06$ mag.
However, the large difference in the $I$-band turnover magnitudes between the blue GCs and red GCs in M87 (0.53 mag) is in strong contrast with the results for Coma galaxies  showing little difference between the two subpopulations 
in this study. At this moment, the reason for this large discrepancy between M87 and Coma galaxies is not known.


Adopting the distance modulus $(m-M)_0=31.09$ (16.5 Mpc) for M87,
\citet{pen09} obtained absolute turnover magnitudes:
$M_I ({\rm max})  = -8.56\pm0.05$ mag for all GCs,  
$M_I ({\rm max})  =-8.85\pm0.06$ mag for the blue GCs,  and 
$M_I ({\rm max})  = -8.32\pm0.09$ mag for the red GCs.
Their value for the blue GCs 
is 0.29 mag brighter than the calibration value adopted in this study, $M_I {\rm (max)} = -8.56\pm0.09$ mag.

\citet{pen09} adopted a distance for the Virgo as the distance to M87, assuming that M87 is at the same distance as the Virgo center.
We checked the distance to M87 to investigate any causes for the difference in the absolute
turnover magnitudes between M87 and Coma galaxies.
We obtained photometry of the resolved stars in M87, 
from deep HST/ACS F606W/F814W images of a field at $6\arcmin$ from M87 available in the HST archive (PID:12989), and used it to estimate the distance to M87 using the tip of the red giant branch (TRGB) method \citep{lee93}. 
We determined the TRGB magnitude to be at $I_{TRGB}=26.85\pm0.05$ mag. 
Applying the TRGB calibration by \citet{riz07} to the TRGB magnitude corrected for foreground extinction ($A_I=0.035$) and systematic photometric offset ($\Delta I=0.042$), we derived a distance to M87, $(m-M)_0=30.90\pm0.13$ and $d= 15.14\pm0.89$ Mpc (Lee \& Jang 2016, in preparation).
This value is $\sim10$\% shorter than the value adopted in \citet{pen09}. 
If this distance is used, 
the turnover magnitude for the blue GCs in M87  in \citet{pen09} will be fainter to $M_I ({\rm max})  =-8.66\pm0.06$ mag. 
This value is consistent with the calibration value adopted in this study, $M_I ({\rm max})  = -8.56\pm0.09$ mag.

To date, GCLFs reaching fainter magnitudes than the turnover magnitudes have been presented only for a small number of galaxies in Coma.
\citet{kav00} presented a GCLF for NGC 4874,  from HST/WFPC2  F606W  and  F814W  images, with $V_0 ({\rm max})  =27.88\pm0.12$ mag and $\sigma =1.49\pm0.12$ for all GCs.
Assuming $(m-M)_0 =30.99$ for Virgo and
$\Delta(m-M)$(Virgo-Coma)$=4.06\pm0.11$, they obtained $(m-M)_0=35.05\pm0.13$ ($d=102\pm6$ Mpc). 
\citet{woo00} published similar results for IC 4051 about two magnitude fainter than NGC 4874: $V_0 ({\rm max})  =27.8\pm0.2$ mag and $\sigma =1.5\pm0.1$.
These results were updated with better data later: 
\citet{har09} suggested a combined GCLF from the HST/WFPC2 data of four giant elliptical galaxies (gEs) in Coma (NGC 4874, 4889, 4926, and IC 4051), 
with a turnover magnitude $V_{\rm max} =27.71\pm0.07$ mag and $\sigma = 1.48$.
For a distance to Coma,  $(m-M)_0 =34.97\pm0.13$ ($d=98.6\pm6.1$ Mpc) based on the recession velocity and the Hubble constant $H_0=72\pm4$ \kmsMpc, they derived $M_V {\rm ({\rm max}) } = -7.32\pm0.13$ mag after correction for extinction ($A_V=0.03$ mag) and K-correction ($K_V = 0.03$ mag \citep{kav00}).
This value is about 0.3 mag fainter than the calibration adopted in this study.

However, the $V$-band turnover magnitudes of  all GCs in the NGC 4921 halo and NGC 4923
($V({\rm max})=27.11\pm0.09$, and $27.18\pm0.09$)  in this study are 0.5 to 0.6 mag brighter than the mean value for Coma galaxies given by \citet{har09}.
For the mean colors, $(V-I)=0.90$ and 1.15 (for the blue and red GCs, respectively), the mean value of the  $V$-band  turnover magnitudes for four Coma galaxies in this study is
$V ({\rm max}) = 27.15\pm0.03$ mag for the blue GCs, 
and  $V ({\rm max})  =27.38\pm0.04 $ mag for the red GCs. 
These values are also 0.3 to 0.5 mag brighter than the \citet{har09} value.
These differences are most likely due to the fact that the images used in this study are deeper and have slightly higher spatial resolutions than those used in the previous study.

Adopting 
$M_I ({\rm max}) = -8.56\pm0.09$ for the metal-poor (blue) GCs,
we obtain distances, from the turnover magnitudes for the blue GCs,
$(m-M)_0=34.80\pm0.10 $ ($d=91\pm4$ Mpc) for NGC 4874, 
and 
$(m-M)_0=34.81\pm0.09$ ($d=92\pm4$ Mpc) for NGC 4889.  
These values are similar to those for NGC 4921 and NGC 4923, showing that all of them are at similar distances and that they are Coma members.
A mean distance for these four Coma galaxies is 
$(m-M)_0=34.80\pm0.09$ ($91\pm4$ Mpc). 
{  Here the error is dominated by the systematic error of the calibration.} 
This value for the distance is slightly smaller than, but consistent with, the value often adopted for Coma, 100 Mpc \citep{car08, pen11}. 
At this distance, one arcmin corresponds to 27 kpc, and one arcsec corresponds to 446 pc.

\subsubsection{The Hubble Constant}

\begin{deluxetable*}{lcc}
\setlength{\tabcolsep}{0.05in}
\tablecaption{Estimation of the Hubble Constant}
\tablewidth{0pt}
\tablehead{ \colhead{Parameter} & \colhead{Value} & \colhead{Remarks}    }
\startdata
Apparent turnover magnitude & $I({\rm max}) = 26.25\pm0.03$ & Mean of four galaxies for the blue GCs\tablenotemark{a} 	\\
Foreground extinction            & $A_I=0.015$ & \citet{sch11} \\
Intrinsic turnover magnitude  & $I_0 = 26.24\pm0.03$ & 	This study\\
Absolute turnover magnitude & $M_I ({\rm max}) = -8.56\pm0.09$ & Metal-poor MW GCs \citep{dic06}\tablenotemark{b}\\
True distance modulus            & $(m-M)_0 = 34.80\pm0.09$ & 	This study\\
Coma distance                        & $d=91\pm4$ Mpc & 	This study \\
Coma velocity w.r.t. CMBR          & $7085\pm95$ \kms & \citet{smi04} \\
Hubble Constant                   &  $H_0 =77.9\pm3.6$ \kmsMpc 
& This study\\
\hline
\enddata
\tablenotetext{a}{Mean for the blue GCs in NGC 4921, NGC 4923, NGC 4874 and NGC 4889.}
\tablenotetext{b}{For the mean color of the  blue GCs, $(V-I)=0.9$. }
\tablenotetext{c}{Error calculations: 
$err(H_0 ) /H_0 = \sqrt{ ((err(v)/v)^2+(err(d)/d)^2) }$,
$x=((m-M)_0+5)/5$, $err(d)/d = ln10~ err(x) = 2.3025~ err(x)$,
$err(x) = err( (m-M)_I + A_I )/5 = 0.2 \sqrt{ err(m)^2+err(M_I )^2 + err(A_I )^2}$. 
}
\label{tab_H0}
\end{deluxetable*}

%
%
%
%

\citet{col96} suggested that the center of the main Coma cluster  is close to  NGC 4874 with $v=6863$ \kms~ and that the velocity dispersion of this region is $\sigma_v= 1082$ \kms. They also suggested that a minor subgroup centered in NGC 4889 ($v=7339$ \kms) with $\sigma_v= 329$ \kms~ is falling into the cluster. The mass of the latter group is estimated to be only 5--10\%, much smaller than the main cluster. Therefore the radial velocity of the main cluster is considered to be that of the NGC 4874 group. 
The Coma velocity with respect to the cosmic background radiation is $7085\pm95$ \kms  \citep{smi04}, which is similar to the value adopted in \citet{kav00}.

From this value and the mean distance to Coma derived in this study, we obtain a value for the Hubble constant,
$H_0=77.9\pm3.6$~\kmsMpc, as summarized in Table 6.
%
This value is consistent with those derived from Type Ia supernovae (SNe Ia) calibrated using Cepheids, $H_0 =74\pm3$ \kmsMpc \citep{rie11, fre12}.
However, it is slightly larger than those based on WMAP and PLANCK observations of the cosmic microwave background radiation \citep{ben13,ben14,pla15} 
or baryonic acoustic oscillations,  
$H_0 =68\pm2$ \kmsMpc, \citep{aub14,ben14}. Note that the most recent value based on PLANK observation is $H_0 =67.8\pm0.9$ \kmsMpc \citep{pla15}.

%


\subsection{Morphological Transformation of NGC 4921} 

The distances to NGC 4921 and NGC 4923  in this study are very similar, and their systemic velocities  ($v=5482$ \kms and 5454 \kms) show little difference. 
 The separation of these two galaxies in the sky is only $2\farcm6$, corresponding to a linear projected distance of 70 kpc at their distance.
This indicates that these two galaxies may be interacting with each other.
The distances to these galaxies are also similar to those of  two main Coma galaxies (NGC 4874 and NGC 4889).
This  shows that NGC 4921 and NGC 4923 are members of the Coma cluster.
The systemic velocity of NGC 4921  is about 1500 \kms~ smaller than that of Coma, 6925 \kms ~(NED). 
Therefore it is concluded that NGC 4921 as well as NGC 4923 may be falling 
toward the Coma center. During this passage NGC 4921 might have lost a significant fraction of its gas due to ram pressure stripping, becoming an anemic spiral galaxy, as suggested by \citet{van76}. 

{  Recently \citet{ken15} provided a  strong evidence for ram pressure acting on NGC 4921, based on the map of HI gas. They found that the distribution of HI gas is asymmetric, being compressed in the northwest side which is directed to the center of Coma, and that the kinematics of the HI gas shows noncircular motion in the northern outer region. They also found that the HI gas is confined inside the stellar disk, and that the HI gas content in this galaxy is only 10\% of a normal galaxy. } 
Thus NGC 4921 may be in the morphological transition from SBab to S0.

We estimated the total number of GCs in NGC 4921 from the GCLF. 
We integrated the bright half of the GCLF for the entire field except for the region of NGC 4923 ($R_{\rm N4923} <1\arcmin$),
and doubled it to get the total number, $N=1400\pm200$.
Then we calculated the specific frequency from this value and the total magnitude of NGC 4921, deriving $S_N = 1.29\pm0.25$. 
This value is higher than the values for typical disk galaxies, and closer to those for massive early-type galaxies \citep{pen08}. As the stellar populations in NGC 4921 evolve, the total luminosity of NGC 4921 gets fainter and its specific frequency increases.
This is consistent with the transitional morphology of NGC 4921. 
 When NGC 4921 becomes an S0, it will be an example
of a massive early-type galaxy formed from a massive progenitor, not via multiple merging.

\subsection{Massive Cluster Formation in Spurs}

The terms 'feathers' \citep{wea70} and 'spurs' \citep{lyn70}  have been used interchangeably in the literature \citep{elm80, kim02, lav06}.
\citet{lav06} summarized their definitions based on observational papers as follows:``Feathers are thin dust lanes or extinction features jutting out at a large angle from the primary dust lanes (the inner side of the arms), and spurs are bright chains of OB associations and HII regions that extend out from the spiral arms into the interarm''.  However, the term 'spurs' is often used to describe dust lane features rather than stellar components.
It is noted that  feathers show few star-forming regions, while spurs are involved often with young star clusters as well as dust lanes \citep{ren13}. 
{  The term `outgrowth' has been also used to represent the spurs \citep{car13,ken15}.

There are numerous studies on the formation of massive star clusters (\citep{zha99,whi10,joh15} and references therein), and some of them addressed the formation of spurs in association with  star clusters.
Numerical simulation studies suggested several mechanisms to  explain the origin of spurs.
First, self-gravity was proposed as a mechanism to form spurs by \citet{cha03}. 
Later Kelvin-Helmholtz instability was introduced to explain wiggle structures in the arms seen in the simulations by \citet{wad04}. Recently high resolution hydrodynamic simulations by \citet{ren13} showed that dense compact spurs were formed by the ejection of the gas from the arms through Kelvin-Helmholtz instability when the velocity gradient is strong (see their Fig. 13), while beads are formed by self-gravity on the spiral arms (see their Figs 12 and 14).
Third, \citet{kim06} found from MHD simulations that spurs are formed via magneto-Jeans instability associated with strong spiral shock waves. 
Fourth, \citet{kim14,kim15} suggested that gaseous feathers (spurs) in grand-design spiral arms can be produced by 
 the wiggle instability of spiral shocks in a galactic disk, and that
 the wiggle instability is produced from the generation of vorticity at a deformed shock front, not from the Kelvin-Helmholtz instability.
Fifth, twisting of the magnetized filaments and the resulting magnetic tension was proposed to explain the features of spurs (outgrowths) in NGC 4921 by \citet{car13} (see their Fig. 14). 
Sixth, ram pressure stripping acting on the magnetized filaments was proposed to describe the properties of the spurs in NGC 4921 by \citet{ken15}.
In this scenario, stars form in the dense gas clouds in the arm where magnetic field has permeated. Then the gas in the arm is stripped by ram pressure, and the dense clouds hosting stars are left off from the arms. The surviving dust clouds which bridge the stars and the arm appear as spurs. 

All these scenarios addressed how the spurs are formed in the presence or absence of star clusters, and they depend on specifically when, where, and how star clusters form at the tip of the spurs.
Among these studies \citet{kim02} and \citet{kim06} suggested from their simulations that bound clouds with  $M\approx 10^7 $ $M_\odot$ can form in the spurs. It is expected that some of these bound clouds may collapse to form massive star clusters with $M\approx 10^5 $ $M_\odot$. However, the spatial resolution in previous simulations was not high enough to investigate the formation of individual star clusters.
Recent studies on NGC 4921 spurs by \citet{car13} and \citet{ken15} addressed  the properties of  spurs, noting the presence of stars and star clusters associated with spurs. However, they did not give any photometric properties of the star clusters associated with the spurs. On the other hand, the study by \citet{tik11} presented the CMD of only the stars, not the star clusters, in the spur region of NGC 4921.

In this study (Figure 5), we presented the CMD of the star clusters in the spur region in comparison with that in Star-forming Region A.
Most of the bright sources detected in these regions are extended star clusters (larger than the GCs).
The brightest clusters in the spur region are much brighter than those in Star-forming Region A, and they are younger than 10 Myr.
The existence of a number of young massive clusters found in the spur region and their masses of $M\approx 10^5 $ $M_\odot$ show indeed that young massive star clusters are forming in the spur region. This is consistent with the prediction for the formation of bound clouds based on theoretical simulations \citep{kim06}. 

What triggered formation of star clusters in the spur region and when did it happen?
Was it triggered by spiral shock waves or by ram pressure?
Were the spur clusters formed before, during, or after the shock due to ram pressure? 
The spatial distributions of the very blue sources in Figure \ref{fig_spat} shows that the concentration in the spur region in the western arm of NGC 4921 is significantly stronger than those in other regions. This indicates that the cluster formation in the spur region might have been triggered by shocks due to ram pressure. 
NGC 4921 provides a precious sample of star clusters that are very useful to answer these questions with theoretical models for the formation of spur clusters in the future. 

\section{SUMMARY AND CONCLUSION}

We analyzed high resolution deep HST/ACS images of NGC 4921 and a part of NGC 4923, after producing the combined images using the AstroDrizzle package.  
Taking advantage of these high resolution drizzled images,  we resolve a significant fraction of GCs in NGC 4921, the brightest spiral galaxy in Coma.  
We presented $VI$ photometry of young star clusters and complexes in the spur and star-forming regions as well as GCs in the entire field. 
In addition, we presented the GCLFs of two cD galaxies in Coma. We combined the GCLFs with the results for NGC 4921 and NGC 4923, deriving a mean distance to Coma.
Primary results are as follows.
\begin{enumerate}

\item
The CMDs of the detected sources confirm the presence of  a rich GC system in NGC 4921 and NGC 4923.
The color distribution of the GCs in NGC 4921 is bimodal.

\item
The GCs in the halo ($70\arcsec<R<160\arcsec$) of NGC 4921 are mostly blue: on average, 0.3 mag bluer ($(V-I)=0.88$) than those in the bulge ($6\arcsec<R\leq 17\arcsec$, $(V-I)=1.15$) of the same galaxy. 

\item We found a number of extended bright star clusters (star complexes) in the spur region of the arms. They are much brighter and bluer than those in Star-forming Region A, being as massive as $M\approx 3 \times 10^5 M_\odot$. This shows that massive star clusters with $M\approx 10^5 M_\odot$ were formed recently in the spur region, consistent with theoretical predictions.

\item 
The turnover magnitudes of the GCLFs for the GCs in NGC 4921 and NGC 4923 are listed in Tables 2 and 3.
The turnover magnitudes of the GCLFs for the blue GCs in the halo of NGC 4921 and NGC 4923 are similar.
$V({\rm max}) =27.11\pm0.09$ mag and $I({\rm max}) =26.21\pm0.11$ for NGC 4921, and
$V({\rm max}) =27.18\pm0.09$ and $I({\rm max}) =26.39\pm0.24$.

\item

We also derived the GCLFs for two cD galaxies, NGC 4874 and NGC 4889 from the HST/ACS images through the same method, obtaining turnover magnitudes 
$I$({\rm max}) $=26.25\pm0.04$ mag and $26.26\pm0.03$ mag for the blue GCs in NGC 4874 and NGC 4889, respectively (see Table 4).   
The turnover magnitudes of the blue GCs show little difference from those of the red GCs, $I$({\rm max}) $=26.25\pm0.04$ mag and $26.26\pm0.03$ mag, respectively.
These values are similar to the results for NGC 4921 and NGC 4923.
The weighted mean values of these four galaxies (NGC 4921, NGC 4923, NGC 4874 and NGC 4889) are  $I$({\rm max}) $=26.25\pm0.03$ for the blue GCs, and 
$I$({\rm max}) $=26.23\pm0.03$ for the red GCs.
For the mean colors, $(V-I)=0.9$ and 1.15 for the blue and red GCs, respectively, corresponding $V$ magnitudes are
$V({\rm max}) =27.15\pm0.03 $ for the blue GCs, and 
$V({\rm max}) =27.38\pm0.04 $ for the red GCs.

%
\item
Adopting 
$M_V ({\rm max}) = -7.66\pm0.09$ mag 
for the metal-poor GCs ([Fe/H]$<-1$) \citep{dic06},
we obtained distances 
$d=88.3\pm5.8$ Mpc for NGC 4921 and 
$d=93.0\pm11.4$ Mpc for NGC 4923. 
Using 
$M_I ({\rm max}) = -8.56\pm0.09$ mag 
for the metal-poor GCs to the $I$-band GCLFs,
we obtained a weighted mean of the distances $d=91\pm4$ Mpc to the four Coma galaxies.

\item
From the mean distance of the four Coma galaxies and the Coma radial velocity with respect to the cosmic background radiation, 
we calculate a value of the Hubble constant to be
$H_0 = 77.9\pm3.6$ \kmsMpc.

\item
We estimated the total number of GCs in NGC 4921,
to be $1400\pm200$ and the specific frequency of GCs to be
$S_N = 1.29\pm0.25$. Considering this and the anemic morphology of NGC 4921, we suggest that NGC 4921 is
in the transition phase to an S0 galaxy.

\end{enumerate}

The authors are grateful to Prof.Woong-Tae Kim for his discussion on the formation of star clusters in spurs,
to the anonymous referee for his/her useful comments, and
to Brian Cho 
for his help in improving the English in the draft.
This work was supported by the National Research Foundation of Korea (NRF) grant
by the Korea Government (MSIP) (No. 2013R1A2A2A05005120).
This paper is based on image data obtained from the Multimission Archive at the Space Telescope Science Institute (MAST).


\clearpage


\begin{thebibliography}{}

\bibitem[Andrade-Santos et al.(2013)]{and13} Andrade-Santos, 
F., Nulsen, P.~E.~J., Kraft, R.~P., et al.\ 2013, \apj, 766, 107 

\bibitem[Aubourg et al.(2014)]{aub14} Aubourg, {\'E}., 
Bailey, S., Bautista, J.~E., et al.\ 2014, arXiv:1411.1074 


\bibitem[Bennett et al.(2013)]{ben13} Bennett, C.~L., Larson, 
D., Weiland, J.~L., et al.\ 2013, \apjs, 208, 20 

\bibitem[Bennett et al.(2014)]{ben14} Bennett, C.~L., Larson, 
D., Weiland, J.~L., \& Hinshaw, G.\ 2014, \apj, 794, 135 

\bibitem[Berkhuijsen et al.(1970)]{ber70} Berkhuijsen, E.~M., 
Haslam, C.~G.~T., \& Salter, C.~J.\ 1970, \nat, 225, 364 

\bibitem[Bernard et al.(2009)]{ber09}Bernard, E.J. et al. 2009, \apj, 699, 1742 

\bibitem[Blanton \& Moustakas(2009)]{bla09} Blanton, M.~R., \& Moustakas, J.\ 2009, \araa, 47, 159 

\bibitem[Brown 
\& Rudnick(2011)]{bro11} Brown, S., \& Rudnick, L.\ 2011, \mnras, 412, 2  

\bibitem[Carlqvist(2013)]{car13} Carlqvist, P.\ 2013, \apss, 343, 689 

 \bibitem[Carter et al.(2008)]{car08} Carter, D., Goudfrooij,  
P., Mobasher, B., et al.\ 2008, \apjs, 176, 424 

\bibitem[Chakrabarti et al.(2003)]{cha03} Chakrabarti, S., 
Laughlin, G., \& Shu, F.~H.\ 2003, \apj, 596, 220  

\bibitem[Colless 
\& Dunn(1996)]{col96} Colless, M., \& Dunn, A.~M.\ 1996, \apj, 458, 435  

 
\bibitem[Davies(1964)]{dav64} Davies, R.~D.\ 1964, \mnras, 
128, 173 

\bibitem[de Vaucouleurs et al.(1991)]{dev91}de Vaucouleurs, G., de Vaucouleurs, A., Corwin, H. G., Jr., et al. 1991, Third
Reference Catalogue of Bright Galaxies (Berlin: Springer) 

\bibitem[Di Criscienzo et al.(2006)]{dic06} Di Criscienzo, 
M., Caputo, F., Marconi, M., \& Musella, I.\ 2006, \mnras, 365, 1357  


\bibitem[Dotter et al.(2008)]{dot08} Dotter, A., Chaboyer, 
B., Jevremovi{\'c}, D., et al.\ 2008, \apjs, 178, 89 



\bibitem[Elmegreen(1980)]{elm80} Elmegreen, D.~M.\ 1980,  
\apj, 242, 528 





\bibitem[Freedman et al.(2012)]{fre12} Freedman, W.~L., 
Madore, B.~F., Scowcroft, V., et al.\ 2012, \apj, 758, 24 



\bibitem[Girardi et 
al.(2000)]{gir00} Girardi, L., Bressan, A., Bertelli, G., \& Chiosi, C.\ 2000, \aaps, 141, 371 
 



\bibitem[Harris et al.(2009)]{har09} Harris, W.~E.,  
Kavelaars, J.~J., Hanes, D.~A., Pritchet, C.~J., 
\& Baum, W.~A.\ 2009, \aj, 137, 3314 







\bibitem[Jang 
\& Lee(2014)]{jan14} Jang, I.~S., \& Lee, M.~G.\ 2014, \apj, 792, 52  

\bibitem[Jang 
\& Lee(2015)]{jan15} Jang, I.~S., \& Lee, M.~G.\ 2015, \apj, 807, 133 

\bibitem[Johnson et al.(2015)]{joh15} Johnson, K.~E., Leroy, 
A.~K., Indebetouw, R., et al.\ 2015, \apj, 806, 35  




\bibitem[Kavelaars et al.(2000)]{kav00} Kavelaars, J.~J., 
Harris, W.~E., Hanes, D.~A., Hesser, J.~E., 
\& Pritchet, C.~J.\ 2000, \apj, 533, 125 

\bibitem[Kenney et al.(2015)]{ken15} Kenney, J.~D.~P., 
Abramson, A., \& Bravo-Alfaro, H.\ 2015, \aj, 150, 59 

\bibitem[Kim et al.(2002)]{kim02} Kim, W.-T., Ostriker, 
E.~C., \& Stone, J.~M.\ 2002, \apj, 581, 1080 

\bibitem[Kim 
\& Ostriker(2006)]{kim06} Kim, W.-T., \& Ostriker, E.~C.\ 2006, \apj, 646, 213 

\bibitem[Kim et al.(2014)]{kim14} Kim, W.-T., Kim, Y., 
\& Kim, J.-G.\ 2014, \apj, 789, 68 

\bibitem[Kim et al.(2015)]{kim15} Kim, Y., Kim, W.-T., 
\& Elmegreen, B.~G.\ 2015, \apj, 809, 33 



\bibitem[Kundu et al.(1999)]{kun99} Kundu, A., Whitmore, 
B.~C., Sparks, W.~B., et al.\ 1999, \apj, 513, 733 

\bibitem[Larsen et al.(2001)]{lar01} Larsen, S.~S., Brodie,  
J.~P., Huchra, J.~P., Forbes, D.~A., 
\& Grillmair, C.~J.\ 2001, \aj, 121, 2974 

\bibitem[La Vigne et al.(2006)]{lav06} La Vigne, M.~A.,  
Vogel, S.~N., \& Ostriker, E.~C.\ 2006, \apj, 650, 818

\bibitem[Lee et al.(1993)]{lee93} Lee, M.~G., Freedman, 
W.~L., \& Madore, B.~F.\ 1993, \apj, 417, 553  






\bibitem[Lee et al.(2015)]{lee15} Lee, G.-H., Hwang, H.~S., 
Lee, M.~G., et al.\ 2015, \apj, 800, 80 

\bibitem[Lim et al.(2013)]{lim13} Lim, S., Hwang, N., 
\& Lee, M.~G.\ 2013, \apj, 766, 20 

\bibitem[Lynds(1970)]{lyn70} Lynds, B.~T.\ 1970, The Spiral 
Structure of our Galaxy, 38, 26 

\bibitem[Muratov 
\& Gnedin(2010)]{mur10} Muratov, A.~L., \& Gnedin, O.~Y.\ 2010, \apj, 718, 1266 


\bibitem[Peng et al.(2008)]{pen08} Peng, E.~W., Jord{\'a}n, 
A., C{\^o}t{\'e}, P., et al.\ 2008, \apj, 681, 197 

\bibitem[Peng et al.(2009)]{pen09} Peng, E.~W., Jord{\'a}n,  
A., Blakeslee, J.~P., et al.\ 2009, \apj, 703, 42 

\bibitem[Peng et al.(2011)]{pen11} Peng, E.~W., Ferguson,  
H.~C., Goudfrooij, P., et al.\ 2011, \apj, 730, 23 



\bibitem[Planck Collaboration et 
al.(2015)]{pla15} Planck Collaboration, Ade, P.~A.~R., Aghanim, N., et al.\ 2015, arXiv:1502.01589 


\bibitem[Rejkuba(2012)]{rej12} Rejkuba, M.\ 2012, \apss, 341, 195 

\bibitem[Renaud et al.(2013)]{ren13} Renaud, F., Bournaud,  
F., Emsellem, E., et al.\ 2013, \mnras, 436, 1836 







\bibitem[Riess et al.(2011)]{rie11} Riess, A.~G., Macri, L., 
Casertano, S., et al.\ 2011, \apj, 730, 119   




\bibitem[Rizzi et al.(2007)]{riz07} Rizzi, L., Tully, R.~B., 
Makarov, D., et al.\ 2007, \apj, 661, 815 






\bibitem[Schlafly 
\& Finkbeiner(2011)]{sch11} Schlafly, E.~F., \& Finkbeiner, D.~P.\ 2011, \apj, 737, 103 

\bibitem[S{\'e}rsic(1963)]{ser63} S{\'e}rsic, J.~L.\ 1963, 
Boletin de la Asociacion Argentina de Astronomia La Plata Argentina, 6, 41  



\bibitem[Sirianni et al.(2005)]{sir05} Sirianni, M., Jee, 
M.~J., Ben{\'{\i}}tez, N., et al.\ 2005, \pasp, 117, 1049 

\bibitem[Smethurst et al.(2015)]{sme15} Smethurst, R.~J., 
Lintott, C.~J., Simmons, B.~D., et al.\ 2015, \mnras, 450, 435 



\bibitem[Smith et al.(2004)]{smi04} Smith, R.~J., Hudson, 
M.~J., Nelan, J.~E., et al.\ 2004, \aj, 128, 1558 


\bibitem[Stetson(1994)]{ste94} Stetson, P.~B.\ 1994, \pasp,  
106, 250  

\bibitem[Struble 
\& Rood(1999)]{str99} Struble, M.~F., \& Rood, H.~J.\ 1999, \apjs, 125, 35  

\bibitem[Tojeiro et al.(2013)]{toj13} Tojeiro, R., Masters, 
K.~L., Richards, J., et al.\ 2013, \mnras, 432, 359 

\bibitem[Tikhonov  
\& Galazutdinova(2011)]{tik11} Tikhonov, N.~A., \& Galazutdinova, O.~A.\ 2011, Astronomy Letters, 37,  766 

\bibitem[van den Bergh(1976)]{van76} van den Bergh, S.\ 1976,  
\apj, 206, 883 

 
\bibitem[Vikhlinin et al.(1997)]{vik97} Vikhlinin, A.,  
Forman, W., \& Jones, C.\ 1997, \apjl, 474, L7 


\bibitem[Wada 
\& Koda(2004)]{wad04} Wada, K., \& Koda, J.\ 2004, \mnras, 349, 270 

\bibitem[Waters et al.(2009)]{wat09} Waters, C.~Z., Zepf, 
S.~E., Lauer, T.~R., \& Baltz, E.~A.\ 2009, \apj, 693, 463 

\bibitem[Weaver(1970)]{wea70} Weaver, H.~F.\ 1970, 
Interstellar Gas Dynamics, 39, 22 

\bibitem[Whitmore et al.(2010)]{whi10} Whitmore, B.~C., 
Chandar, R., Schweizer, F., et al.\ 2010, \aj, 140, 75 
	
	
\bibitem[Woodworth 
\& Harris(2000)]{woo00} Woodworth, S.~C., \& Harris, W.~E.\ 2000, \aj, 119, 2699  

\bibitem[Zhang 
\& Fall(1999)]{zha99} Zhang, Q., \& Fall, S.~M.\ 1999, \apjl, 527, L81 

%
%
\end{thebibliography}
\end{document}